\documentclass[12pt,]{article}
\usepackage{lmodern}
\usepackage{graphicx}
\usepackage{natbib}
\usepackage{subcaption}
\usepackage{amssymb,amsmath}
\usepackage[margin=1.1in]{geometry}
\usepackage{hyperref}
\hypersetup{unicode=true,
            pdftitle={Combining social media and survey data to nowcast migrant stocks in the US},
            pdfauthor={Monica Alexander; Kivan Polimis; Emilio Zagheni},
            pdfborder={0 0 0},
            breaklinks=true}
\usepackage{graphicx,grffile}
\makeatletter
\def\maxwidth{\ifdim\Gin@nat@width>\linewidth\linewidth\else\Gin@nat@width\fi}
\def\maxheight{\ifdim\Gin@nat@height>\textheight\textheight\else\Gin@nat@height\fi}
\makeatother
\setkeys{Gin}{width=\maxwidth,height=\maxheight,keepaspectratio}
\IfFileExists{parskip.sty}{%
\usepackage{parskip}
}{% else
\setlength{\parindent}{0pt}
\setlength{\parskip}{6pt plus 2pt minus 1pt}
}
\providecommand{\tightlist}{%
  \setlength{\itemsep}{0pt}\setlength{\parskip}{0pt}}
\ifx\paragraph\undefined\else
\let\oldparagraph\paragraph
\renewcommand{\paragraph}[1]{\oldparagraph{#1}\mbox{}}
\fi
\ifx\subparagraph\undefined\else
\let\oldsubparagraph\subparagraph
\renewcommand{\subparagraph}[1]{\oldsubparagraph{#1}\mbox{}}
\fi

\let\rmarkdownfootnote\footnote%
\def\footnote{\protect\rmarkdownfootnote}

\usepackage{titling}

  \title{\sc Combining social media and survey data to nowcast migrant stocks in the United States}
    \pretitle{\vspace{\droptitle}\centering\huge}
  \posttitle{\par}
    \author{Monica Alexander\footnote{University of Toronto
 \texttt{monica.alexander@utoronto.ca}.} \\ Kivan Polimis\footnote{University of Washington} \\ Emilio Zagheni\footnote{Max Planck Institute for Demographic Research}}
    \preauthor{\centering\large\emph}
  \postauthor{\par}
    \date{}
    \predate{}\postdate{}
  
\usepackage{setspace}
\begin{document}
\maketitle
\begin{abstract}
Measuring and forecasting migration patterns, and how they change over time, has important implications for understanding broader population trends, for  designing policy effectively and for allocating resources. However, data on migration and mobility are often lacking, and those that do exist are not available in a timely manner. Social media data offer new opportunities to provide more up-to-date demographic estimates and to complement more traditional data sources. Facebook, for example, can be thought of as a large digital census that is regularly updated. However, its users are not representative of the underlying population. This paper proposes a statistical framework to combine social media data with traditional survey data to produce timely `nowcasts' of migrant stocks by state in the United States. The model incorporates bias adjustment of the Facebook data, and a pooled principal component time series approach, to account for correlations across age, time and space. We illustrate the results for migrants from Mexico, India and Germany, and show that the model outperforms alternatives that rely solely on either social media or survey data. 
\end{abstract}

%%%%%%%%%%%%%%%%%%%%%%%%%%%%%%%%%%%%%%%%%%%%%%%%%%%%%%%%%
\newpage
\section{Introduction}\label{introduction-and-motivation}

Accurate, reliable and timely estimates of migration indicators, such as flows and stocks, are crucial for understanding population dynamics and demographic change, for designing effective economic, social and health policies, and for supporting migrants and their families. However, data on migration from traditional sources, such as censuses,  surveys or administrative registers, are often insufficient.  Even when these sources exist, the data available may lack the granularity of information required to understand migration trends, or are not released in a manner that is timely enough to monitor changes in  trends. 
%In the United States, one of the most authoritative sources on international migrant stocks is the American Community Survey (ACS); however, there is usually at least a one-year delay in the release of the public-use data.
As migration flows can change substantially over a short period of time --- such as in response to a natural disaster or war and conflict --- relying on out-dated data is often not sufficient. 

Timely and reliable information about migration stocks is important not only to understand migration patterns. It is key also to monitor fertility, population health and mortality. Even when accurate data on births and deaths exist, often demographers are faced with large uncertainty in population counts, which, in various disaggregations, form the denominators for standard demographic rates. Most of the uncertainty is driven by lack of appropriate information on how migration stocks change over time and space.

As a consequence of data availability issues, we need to consider how  non-traditional data can be leveraged to complement existing sources in order to improve estimates and predictions of migration indicators over time. Previous work has explored the use of data such as call detail records \citep{blumenstock2012inferring,pestre2020abcde}, air traffic data \citep{gabrielli2019dissecting}, tax file records \citep{engels1981measuring} and other sources like billing addresses or school enrollment \citep{foulkes2008using} to estimate migration. Additionally, an increasingly large body of work has investigated the use of social media data, from websites such as Twitter \citep{zagheni2014inferring}, Facebook \citep{zagheni2017} and LinkedIn \citep{rodriguez2014migration}. Provided that the data can be obtained in a reliable, timely, and ethical way, information about the users of social media websites is potentially an incredibly rich demographic data source. 
%The users of a social media platform could be thought of as a population, with new accounts like `births', deleted accounts like `deaths', and a whole series of movements in between. 
Data on these populations are essentially collected in real time, and while individual-level information is usually restricted, many of the social media websites provide a certain amount of aggregate-level information through their advertising platforms \citep{cesare2018promises}. In particular, Facebook's Advertising platform allows information to be extracted on the relative size of groups by key demographics such as age, sex, location of residence and country of origin, thereby potentially acting as a measure of the relative size of migrant groups in a particular country. 

While these data have clear potential for use in demographic research, with respect to timeliness and the size of the sample being considered, there are some notable issues that need to be overcome. In particular, for any given population subgroup of interest, the corresponding users of Facebook or any other social media platform are unlikely to be a representative sample. An additional challenge is to use these data in a way that meaningfully combines  new `signal' or information about migration trends with existing knowledge on probable migration trends from historical data sources. 

This paper proposes a statistical framework to combine social media data from Facebook, with traditional survey data from the ACS, in order to produce timely `nowcasts' of migrant stocks by state in the United States. The framework consists of a Bayesian hierarchical model which incorporates bias adjustment of the Facebook data, a demographic time series approach to account for historical past trends, and a geographic pooling component which allows information about the age structure of migrants to be shared across space. The model also accounts for the different types of uncertainty that are likely to be present in Facebook and traditional survey data. The resulting model produces estimates and short-term projections of migrant stocks by US state of destination  and country of origin, and is shown to out-perform valid modeling alternatives. 

The remainder of the paper is structured as follows. First, we briefly discuss previous demographic research which incorporates social media. Then we outline the data sources used, and in particular how the Facebook data were collected. Section 4 discusses the model set-up, assumptions and computation. We then present results for Mexican, Indian and German migrants by US state, and validate model performance against reasonable alternatives. Finally, the strengths and limitations of the model are discussed, together with avenues for future research.

\section{Background}

%EMILIO PLEASE ADD ANYTHING RELEVANT/MORE UP TO DATE???

The lack of good-quality data on migration is a global problem, with data sparsity issues prevalent in both developed and developing countries \citep{landau2017}. This has prompted scholars to investigate the use of other types of data to monitor migration trends. In particular, with the rise of social media use around the world, new data that have potential for demographic research have emerged. 

Scholars began using social media and web data to estimate and track demographic indicators over time in the early 2010s. The earliest papers illustrated how geo-located data from email services and web-based applications such as Twitter, Google Latitude, Foursquare or Yahoo! can be used \citep{ferrari2011,noulas2011empirical,zagheni2012you}. Initial research focused on evaluating spatial mobility of populations at a city or regional level. For example, \cite{ferrari2011} used Twitter data to study patterns of urban movement in New York. In the first effort to tackle global trends, \cite{zagheni2012you} linked the geographic locations of IP addresses of Yahoo! emails to the user's self-reported demographic data to estimate age- and sex-specific migration flows in a large number of countries around the world. 

Recent efforts have focused on using data from social media and networking websites such as Twitter, LinkedIn and, more recently, Facebook and Instagram. These websites provide public access to an Application Programming Interface (API), which makes it possible to send requests and receive responses from these websites for data like tweet hashtag counts, the number of jobs in a certain industry, or number of cell phone users in a particular area. Researchers have utilized these APIs to extract publicly available demographic and location data for use in social research, in particular to study outcomes such as migration \citep{yildiz2017,zagheni2017}, fertility \citep{rampazzo2018mater}, gender equality \citep{fatehkia2018using,garcia2018analyzing}, and health \citep{araujo2017using}. For instance, \cite{garcia2018analyzing} used Facebook data to create an index of the internet gender divide in 217 countries, showing that this indicator encapsulated gender equality indices in education, health and economic opportunity. \cite{yildiz2017} used a combination of geo-located tweets and image recognition software to obtain estimates of internal migration in England. 

In work relevant to this paper, \cite{zagheni2017} presented a proof of concept for estimating migration stocks in the United State by age, sex and state, using Facebook's Advertising Platform. More recently,  \cite{alexander2019impact} used the same type of data to track changes in migrants over time, in the context of estimating out-migration from Puerto Rico following Hurricane Maria in September 2017. 

The main gap in the literature is related to the lack of a suitable statistical model for combining `traditional' data sources on migrants --- in the form of censuses, nationally representative surveys, or other vital statistics --- with migration information from social media data. The goal of this paper is thus to develop a probabilistic framework that allows representative and historical time series to be combined, in a sound statistical framework, with non-representative -- but timely -- sources from social media.

%%%%%%%%%%%%%%%%%%%%%%%%%%%%%%%%%%
\section{Data}\label{data}

\subsection {Facebook Advertising data}
Facebook for Business has developed a targeted advertising platform, called Ads Manager that provides a graphical user interface to allow advertisers to micro-target specific audiences. Demographic characteristics that can be targeted include information directly reported by Facebook users, such as age or sex, and information indirectly inferred from using Facebook's platform or affiliated websites, such as location and behavioral interests. Before launching an advertisement, an advertiser can select a variety of characteristics (e.g., Australians living in California, who are female, and aged 30-35) and get an estimate of the `potential reach' (monthly active users) to this subgroup. These estimates can be obtained, in a programmatic way, for a variety of different expatriate (`expat') groups by age, sex, and education. 

We use the estimates of potential reach by expat group, age and sex to track sizes of migration stocks over time. These estimates can be obtained before the launch of an advertisement, and as such are obtained free of charge. We use the Ads Manager backend application, Facebook's Marketing API, to extract estimates of potential reach over time programmatically with the Python module pySocialWatcher \citep{araujo2017facebook}. With pySocialWatcher, we collected data across 11 age groups (10 UN age groups from 15-19 to 60-65; an 11th group for the entire available Facebook population of 13-65 was also used), three gender groups (female, male, and total population) and multiple education categories. Data was collected using Amazon Web Services (AWS) EC2 Instance servers.

As part of a broader project on using social media in demographic research, we started data collection in January 2017. For each wave of data collection we obtained state-level estimates of all Facebook users (by age, sex, and gender) as well as state-level estimates of 90 expat groups. We have been collecting a new wave of data every 2-3 months. \footnote{The waves used in this paper are:
 Wave 1: January 2017; 
 Wave 2: April 2017; 
 Wave 3: June 2017; 
 Wave 4: October 2017; 
 Wave 5: January 2018; 
 Wave 6: March 2018}

\subsection{American Community Survey}

The American Community Survey (ACS) is an annual survey of the U.S. Census Bureau, designed to supplement the decennial census. Based on the long-form version of the census, the ACS collects information on topics including population, housing, employment and education from a nationally representative sample. Data on migrant stocks can be readily obtained from the ACS. In particular, in every year of the ACS, the survey has contained a question asking the birthplace of the person; if it is inside the United States, the state is recorded, and if it is outside the United States, the country is recorded. This birthplace variable is recorded as a three digit code to indicate the US state or country of birth. In addition to the birthplace variable, the ACS has information on current state of residence. Thus, we can tabulate the number of migrants from a particular country living in a particular state by looking at the combination of these two variables. From a modeling perspective, we are interested in the proportion of migrants from a particular origin of the total population by five-year age groups (\(15-19, 20-24, \dots, 50-54\)) in each state. 

We calculated the migrant stock proportions using the 1-year ACS for each year between 2001 and 2017 using micro-data available through the Integrated Public Use Microdata (IPUMS) US project \citep{ruggles2000}. Standard errors around the calculated proportions based on sampling variation were calculated based on ACS accuracy guidelines \citep{uscbacs} and using the Delta method. 

\section{Model}\label{model}

We are considering two data sources of migration
trends in the US: data from Facebook's Advertising Platform, and the
ACS. The overall goal of the modeling strategy is to combine information
from both these sources to produce estimates of current and future
migrant stocks. To do this, the model should have three main
characteristics. Firstly, we want to adjust for biases in Facebook data
to effectively use up-to-date information on migration patterns from
this source. Secondly, we want to be able to incorporate longer time
series of information from the ACS. Finally, the data should be combined
in a probabilistic way, in order to objectively weigh information from
both sources. We propose a Bayesian hierarchical model which achieves
these goals. In this section we describe the model in detail.

For a particular migrant group, define \(\mu_{xts}\) to be the (`true')
proportion of migrants of the total populations in age group \(x\) at
time \(t\) and in state \(s\). This quantity \(\mu_{xts}\) is the main
parameter of interest to be estimated. We have observations of this
proportion, which will be denoted \(p_{xts}\). The observed proportions
are either from Facebook (\(p_{xts}^{FB}\)) or from the ACS (\(p^{ACS}_{xts}\)).
The \(p_{xts}\)s are observed, and it is assumed that these are somehow
related to the underlying true proportions, \(\mu_{xts}\), with some associated
error. We use the term `true' in a statistical sense, referring to a latent variable of interest.

\subsection{Facebook bias adjustment}\label{facebook-bias-adjustment}

\label{model_fb}

The first goal is to adjust the Facebook data to account for the
non-representativeness of the Facebook user population. Previous
research from \cite{zagheni2017} showed that, while the
bias in the Facebook migrant data is substantial, it is also relatively
systematic by age and migrant group and can be modelled.

Following their approach, we introduce a regression model which relates
the proportions of migrants in Facbook, \(p_{xts}^{FB}\), to the
proportions in the ACS in a similar time period, \(p_{xts}^{ACS}\), plus
a series of age and state variables. In particular, for a particular
migrant group, express \(p_{xts}^{ACS}\), on a log scale, as

\begin{equation}
\label{eqn_fb}
\log p_{xts}^{ACS} = \alpha_0 + \alpha_1 \log p_{xts}^{FB} + \mathbf{\beta X} + \varepsilon_{FB}
\end{equation}

where \(\mathbf{X}\) is a covariate matrix containing an indicator
variable for each age group (\(15-19, 20-24, \dots, 50-54\)) and each of
the 50 states plus Washington D.C. This means that we estimate a fixed effect for each age group and state. In addition, we assume that the error is i.i.d and
that

\[
\varepsilon_{FB} \sim N(0, \sigma^2_{FB}). 
\]

Estimates of the coefficients \(\alpha_0\), \(\alpha_1\) and the vector
of \(\beta\)'s are obtained using the first wave of the Facebook data
and the 2016 ACS data. Once obtained, these coefficient estimates are
then used to adjust subsequent waves of Facebook data, i.e.~we calculate

\[
\log p_{xts}^{*} = \hat{\alpha}_0 + \hat{\alpha}_1 \log p_{xts}^{FB} + \mathbf{\hat{\beta} X}
\] where \(\log p_{xts}^{*}\) is a `bias-adjusted' version of the
Facebook data. This is taken to be our `best guess' of what the migrant
stocks in group \(xts\) are, based on the Facebook data alone. Note that
an estimate of \(\sigma^2_{FB}\) is also obtained, that is, the variance of the error terms, which becomes
important in the final model (see section \ref{BIAT} below).

\subsection{Time series modeling of ACS using principal
components}\label{time-series-modeling-of-acs-using-principal-components}

\label{model_acs}

In addition to using data from Facebook, we also want to incorporate the
relatively long historical time series of information on migrant stocks
obtained from the ACS. A reasonable short term forecast based on ACS
should model historical trends and project them forward.

There are many different time series models that could be used in this
context. Perhaps the simplest approach would be to project forward a moving average of the time series for each age group and state combination. Alternatively, we could use a classical Box-Jenkins approach and
model the time series of migrant stocks in each age group and state
separately using an appropriately specified ARIMA model. However, these methods would not place any constraints on the age structure of migration. Given
this demographic context, we expect that the age distribution of migration
displays strong patterns and changes in a relatively regular way over
time. This is because of regularities in the age at migration as well as historical trends which include different waves of migrants, who also age over time. As such, we chose to incorporate this prior knowledge into our
model through a principal components approach.

Principal component-based models have a long history in demographic
modeling, with the most well known example being the Lee-Carter
mortality model \citep{lee1992}. The idea is that a set of age-specific demographic
rates observed over time can be expressed as a combination of a series
of characteristic patterns (or principal components). The Lee-Carter
approach uses the mean age-specific mortality schedule and first
principal component, which is interpreted as age-specific contributions
to mortality change over time. This model can easily be extended
to include higher-order principal components, which various researchers
have done.

Apart from the Lee-Carter model and variants (e.g. \cite{li2004using}, \cite{lee2000lee}, \cite{renshaw2006cohort}), principal
component models have been recently used to estimate and forecast
mortality (e.g. \cite{alexander2017flexible}), fertility (e.g. \cite{schmertmann2014bayesian}) and overall population (\cite{wisniowski2015bayesian}). Here, we extend this idea to
parsimoniously estimate and project migration stocks by age and state.

\subsubsection{Model overview}\label{model-overview}

Age-specific migration schedules are decomposed into
independent age and time components. The time component is then projected forward as a time series, taking auto-correlated error into account. We propose a log-linear model for
\(p_{xts}\):

\begin{equation}
\label{eqn_acs}
\log p_{xts}^{ACS} = \beta_{ts,1} Z_{x,1} + \beta_{ts,2} Z_{x,2} + \varepsilon_{xts}
\end{equation}

where \(Z_{x,1}\) and \(Z_{x,2}\) are the first and second `principal
components', \(\beta_{ts,1}\) and \(\beta_{ts,2}\) are state and
time-specific coefficients, to be estimated, and \(\varepsilon_{xts}\)
is an error term. The principal components are obtained via Singular Value
Decomposition (SVD), as outlined in the next section. To obtain
estimates of \(\beta_{ts,1}\) and \(\beta_{ts,2}\), we impose some
smoothing over time and pooling of information across space, as outlined
in section \ref{hierarchical}. Finally, as discussed in Section
\ref{autocorrelation} we place a time series model on the error term,
\(\varepsilon_{xts}\), to account for auto-correlation.

\subsubsection{Obtaining the principal
components}\label{obtaining-the-principal-components}

The principal component terms \(Z_{x,1}\) and \(Z_{x,2}\) aim to capture
the main sources of systematic variation in migration patterns across
age. They are obtained by first creating a matrix of (logged) historical
age-specific migration schedules based on ACS data from 2001 to 2016.
Singular Value Decomposition (SVD) is then performed on this matrix to
obtain principal components of the age-specific migration. In
particular, let \(\bf{X}\) be a \(N \times G\) matrix of log-migration
stock rates, where \(N\) is the number of state-years and \(G\) is the
number of age-groups. In this case, we had \(N = 51\) states + DC
\(\times 16\) years \(= 816\) observations of \(G = 9\) age-groups
(\(15-19, 20-24, \dots, 50-54\)). The SVD of \(\bf X\) is

\begin{equation}
\bf X = \bf{UDV'},
\end{equation}

where \(\bf U\) is a \(N \times N\) matrix, \(\bf D\) is a
\(N \times G\) matrix and \(\bf V\) is a \(G \times G\) matrix. The
first two columns of \(\bf V\) (the first two right-singular values of
\(\bf X\)) are \(Z_{x,1}\) and \(Z_{x,2}\).

\begin{figure}
\centering
\includegraphics[width=1\textwidth]{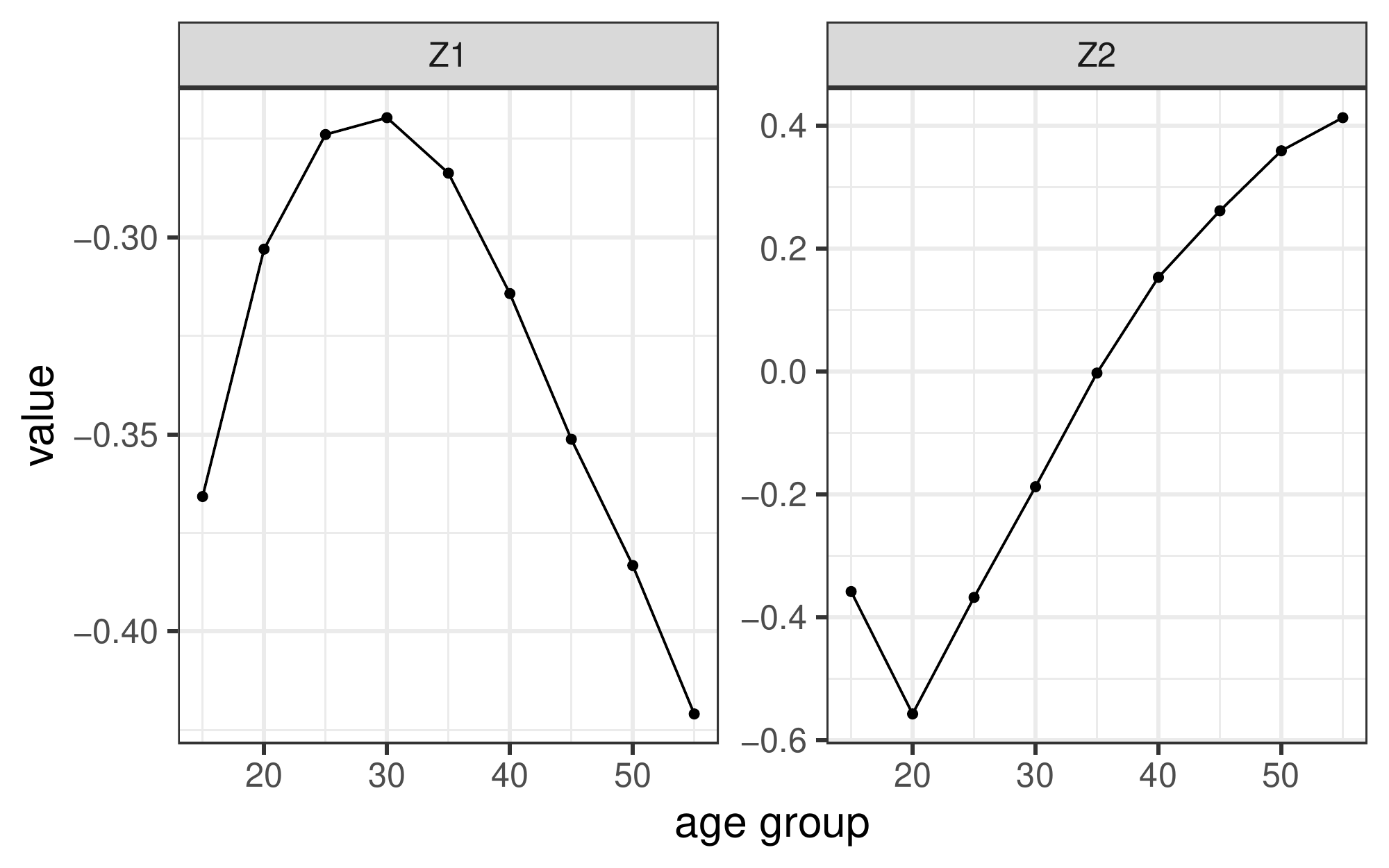}
\caption{Principal Components for Mexico}
\label{pcs}
\end{figure}

For example, Fig.~\ref{pcs} shows the resulting \(Z_{.,1}\) and
\(Z_{.,2}\) for the Mexican migrant group in the US. These were obtained
via the following steps:

\begin{enumerate}
\def\labelenumi{\arabic{enumi}.}
\tightlist
\item
  Calculate \(p_{xts}^{ACS}\), i.e.~the proportion of migrants in age
  group \(x\), year \(t\) and state \(s\) for each age group, year and
  state in ACS 2001-2016.
\item
  Create \(\bf{X}\) where each element is \(\log p_{xts}^{ACS}\), every
  row is a state-year and every column is an age group.
\item
  Perform SVD on \(\bf X\)
  and extract the first two columns\footnote{We used the `svd' function in R.} of \(\bf V\).
\end{enumerate}

The principal components shown in Fig.~\ref{pcs} can be interpreted as a
baseline migration age schedule (\(Z_{.,1}\)) and age-specific
contributions to change over time (\(Z_{.,2}\)). In the model, the
coefficient on \(Z_{.,1}\) (\(\beta_{ts,1}\)) moves the overall level of
Mexican migrants up or down, depending on the year and state. The
coefficient on \(Z_{.,2}\) allows the age distribution to shift to older
or younger ages. For \(Z_{.,2}\), the sign changes from negative to
positive at age 35. This means that the larger and more positive the value of
\(\beta_{ts,2}\), the older the migrant age
distribution.\footnote{Note that the interpretation of $Z_{.,1}$ and $Z_{.,2}$ is similar to the interpretation of the $a_x$ and $b_x$ terms in the usual Lee-Carter model.}

\subsubsection{Sharing information across time and
space}\label{sharing-information-across-time-and-space}

\label{hierarchical} The model specified in Eqn.~\ref{eqn_acs} requires
the estimation of two coefficients, \(\beta_{ts,1}\) and
\(\beta_{ts,2}\) for each time \(t\) and state \(s\). One option would
be to estimate each of these coefficients separately for every year and state. However, we would
like to incorporate the knowledge that trends in migration over time are
likely to exhibit relatively regular patterns. In addition,
for the coefficient on the second principal component --- which allows
for the age distribution of migrants to shift to the left or right, we
would like to share information about the patterns in migration across
geographic space.

The coefficient on the first principal component, \(\beta_{ts,1}\), is
modeled as a random walk, i.e.

\[
\beta_{ts,1} \sim N(\beta_{t-1s,1}, \sigma_{\beta_1}^2)
\]

This allows for information about the level of migration within each
state to be smoothed over time. The random walk structure allows for the
estimate in the current time period, \(\beta_{ts,1}\), to be partially
informed by the previous period.

For the coefficient on the second principal component, we place the
following hierarchical structure on the \(\beta\)'s:

\[
\beta_{ts,2}  \sim N(\Phi_{t}, \sigma_{\beta}^2)
\] \[
\Phi_{t} \sim N(\Phi_{t-1}, \sigma_{\Phi}^2)
\]

where \(p = 1,2\). The \(\Phi_{t}\) term represents  essentially a
national mean; as such the \(\beta_{ts,2}\)'s are a draw from a national
distribution with some mean and variance. In this way, information about
how the age distribution is ageing over time is shared across states.
The more information about migration there is available for a particular
state (i.e., the larger the migrant population), the less the estimate
of \(\beta_{ts,2}\) is influenced by the overall mean. Conversely,
states with smaller migrant populations where the trends over time are
less clear from the data are partially informed by patterns in larger
states.

Note that the geographical hierarchical structure is not present on the
first coefficient, as this represents an overall level of migration.
Pooling information across space about the level of migration would
artificially increase migrant proportions in smaller states.

\subsubsection{Auto-correlated error}\label{autocorrelated-error}

\label{autocorrelation}

The final piece of the time series model is the error term
\(\varepsilon_{xts}\). This term is included in the model to allow for
extra variation in migration age schedules that is not otherwise picked
up by the principal components. We expect  the extra variation to be
autocorrelated, and as such we model the error term as an AR(1) process:

\[
\varepsilon_{xts} \sim N(\rho_{xs}\varepsilon_{xt-1s}, \sigma_{\varepsilon}^2)
\] where \(\rho_{xs} \in [0,1]\).

\subsubsection{Projection}\label{projection}

The model described above is fit to ACS data from 2001-2016. However,
estimates in more recent years can easily be obtained by projecting the
time series aspects of this model forward. In particular, for time
\(t+1\):

\begin{itemize}
\tightlist
\item
  Obtain an estimate for \(\beta_{t+1s,1}\) from
  \(\beta_{t+1s,1} \sim N(\beta_{ts,1}, \sigma^2_{\beta})\).
\item
  Obtain an estimate for \(\beta_{t+1s,2}\) from
  \(\beta_{t+1s} \sim N(\Phi_{t+1}, \sigma^2_{\Phi})\) and
  \(\Phi_{t+1} \sim N(\Phi_{t}, \sigma_{\Phi}^2)\).
\item
  Obtain an estimate for \(\varepsilon_{xt+1s}\) from
  \(\varepsilon_{xt+1s} \sim N(\rho_{xs}\varepsilon_{xts}, \sigma_{\varepsilon}^2).\)
\item
  Calculate \(\log p_{xt+1s}^{ACS}\) based on Eqn.~\ref{eqn_acs}.
\end{itemize}

\subsection{Bringing it all together}\label{bringing-it-all-together}

\label{BIAT}

Sections \ref{model_fb} and \ref{model_acs} described two ways
to obtain current `nowcasts' of migrant stocks. One option would be to
take the most recent data obtained from Facebook, adjust using the
bias-adjustment model, and take the resulting estimate as our nowcast.
Another option would be to project forward the ACS model to the time
period of interest. Ideally, we would like to incorporate both sources
into our final estimate. Perhaps an option would be to just take
an average of the two resulting estimates. However, we would like
to  weigh the estimates from both sources more objectively, taking different
sorts of uncertainty into consideration.

\begin{figure}
\centering
\includegraphics[width=1\textwidth]{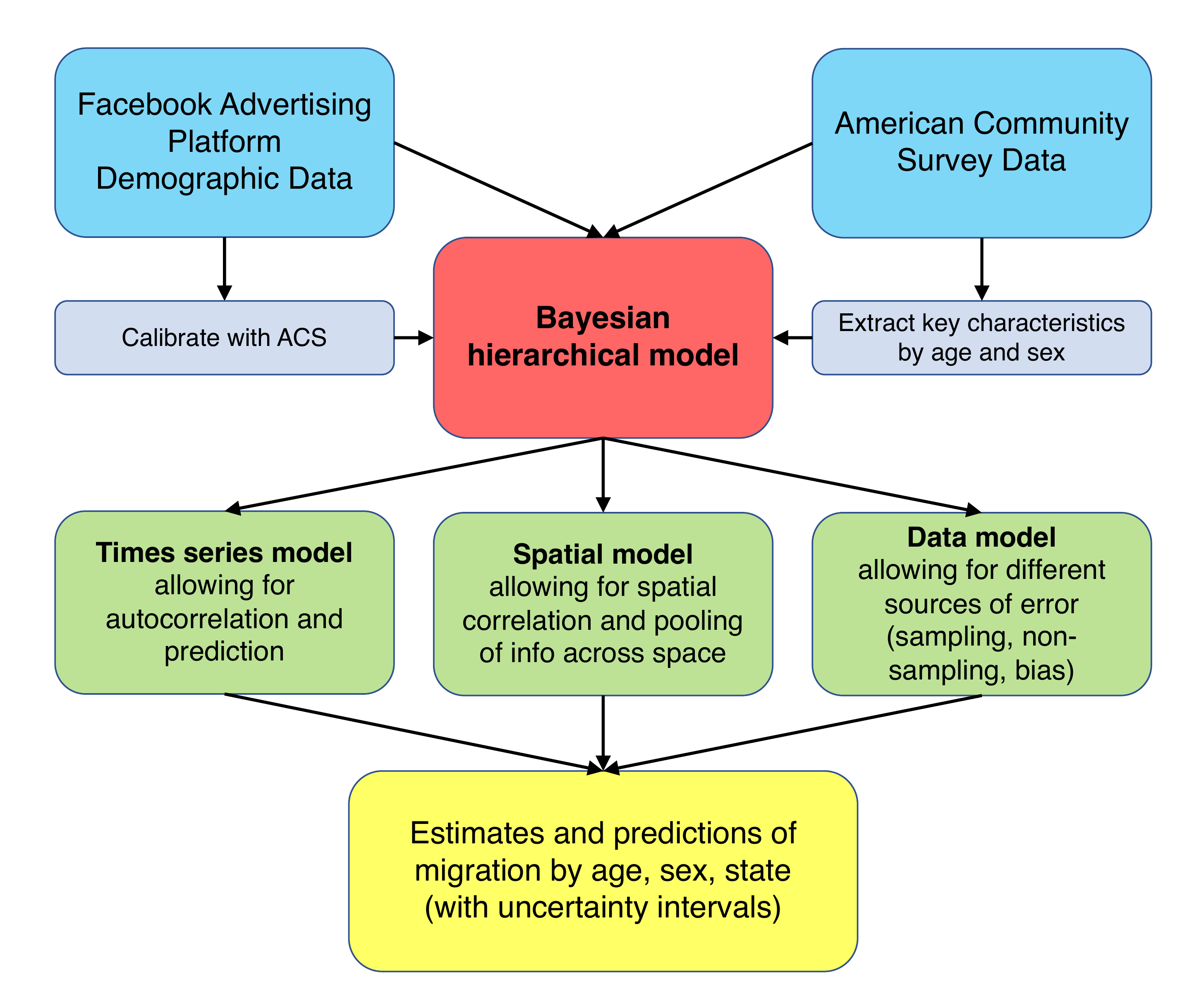}
\caption{Modeling framework}
\label{fig_flow}
\end{figure}

Our solution is to combine both models into one framework, and use
the results from both methods as data points for our `best estimate'
nowcast. This is illustrated in Fig.~\ref{fig_flow}. Facebook inputs are
calibrated with the ACS via the adjustment model. ACS data are used to
obtained principal components based on past migration data. The modeling
structure allows for information exchange over time and across
geographic space. The key piece of the combined model, which has yet to
be explained, is the data model (or likelihood), which allows data from
the different sources to have different associated error.

\subsubsection{Data model}\label{data-model}

As outlined above, we observe migrant proportions \(p_{xts}\) from
either Facebook or the ACS. The data model assumes \[
\log p_{xts} \sim N(\log \mu_{xts}, \sigma_p^2) 
\] i.e. the log of the observed proportion is assumed to have mean $\log \mu_{xts}$ and variance \(\sigma_p^2\),  where \(\sigma_p^2\) depends on the data source:

\[
\sigma_p^2 = 
\begin{cases}
    \sigma_s^2,& \text{if ACS} \\
    \sigma_s^2 + \sigma_{FB}^2 + \sigma_{ns}^2, & \text{if Facebook}
\end{cases} \label{eq:sigma}\\
\] Here, \(\sigma^2_s\) refers to sampling error, and is assumed to be
present in both ACS and Facebook data. For the ACS data, sampling errors
are calculated based on guidelines from the \cite{uscbacs}. For Facebook data, the sampling error is
calculated assuming the binomial approximation to the Normal distribution
and calculating

\[
\sigma^2_s = \frac{p_{xts} \cdot (1-p_{xts})}{N_{xts}^{FB}}
\]

where \(N_{xts}^{FB}\) is the total size of the Facebook population in
subgroup \(x,t,s\).

For the Facebook data there are two additional error terms.
\(\sigma_{FB}^2\) refers to the error associated with our
bias-adjustment model (Eqn. \ref{eqn_fb}) and is estimated within this
model. This captures the fact that our adjustment model is imperfect and
that extra variation remains. Additionally, we allow for a non-sampling
error with \(\sigma_{ns}^2\), which aims at capturing additional uncertainty like variation in
the way potential reach is estimated across waves.

For a given population size, the sampling error is going to be of similar
size for ACS and Facebook data. As such, the error term associated with
the Facebook data, which is the sum of three terms, will always be bigger
than for ACS. In practice, this means that estimates from the model will
follow (i.e.~give more weight to) the ACS data.

\subsubsection{Summary of full model}\label{summary-of-full-model}

The full model is summarized below. Equation 4 is the data model.
Equations 5-9 relate to the ACS time series model. Equation 12 is
related to the Facebook regression model. Equations 10 and 11 allow the
observation of the proportion of interest to come from a different
source (Facebook or ACS), which has a different associated variance.
Note that \(\mu_{xts}\) is estimated on a yearly basis, but it is
assumed that \(j\) waves of Facebook data are collected within any one
year.

\begin{eqnarray}
\log p_{xts} &\sim& N(\log \mu_{xts}, \sigma^2) \label{eq:log} \\
\log \mu_{xts} &=& \beta_{ts,1} Z_{x,1} + \beta_{ts,2} Z_{x,2} + \varepsilon_{xts} \label{eq:prcomp} \\
\beta_{ts,1} &\sim& N(\beta_{t-1s,1}, \sigma_{\beta_1}^2)\\
\beta_{ts,2} &\sim& N(\Phi_{t,2}, \sigma_{\beta}^2)\\
\Phi_{t} &\sim& N(\Phi_{t-1}, \sigma_{\Phi}^2)\\
\varepsilon_{xts} &\sim& N(\rho_{xs}\varepsilon_{xt-1s}, \sigma_{\varepsilon}^2)\\
p_{xts} &=&
\begin{cases}
    p_{xts}^{ACS}, & \text{if } 2001\leq t \leq 2016\\
    p_{xtsj}^*,  & \text{if } t\geq2017 
\end{cases} \label{eq:proportion} \\
\sigma^2 &=& 
\begin{cases}
    \sigma_s^2,& \text{if ACS}\\
    \sigma_s^2 + \sigma_{FB}^2 + \sigma_{ns}^2, & \text{if Facebook} \\
\end{cases}\\
p_{xtsj}^* &\sim& N ({\alpha_0} + {\alpha_1} \cdot p_{xtsj}^{\text{ Facebook}} + X{\Gamma}, \sigma^2_{FB})
\end{eqnarray}

\subsubsection{Priors}\label{priors}
Weakly-informative priors were placed on the coefficients in the Facebook bias-adjustment model, as well as the principal component coefficients in the initial periods:
\begin{eqnarray*}
\alpha_0 &\sim& N(0, 100)\\
\alpha_1 &\sim& N(0, 100)\\
\Gamma_0 &\sim& N(0, 100)\\
\beta_{1,s,1} &\sim& N(0, 100)\\
\Phi_{1} &\sim& N(0, 100)
\end{eqnarray*}

In addition, we put weakly-informative half-Normal priors on the two standard deviation terms to be estimated:
\begin{eqnarray*}
\sigma_{FB} &\sim& N_+(0, 1)\\
\sigma_{ns} &\sim& N_+(0, 1).
\end{eqnarray*}

\subsubsection{Computation}\label{computation}
The model was fitted in a Bayesian framework using the statistical software R. Samples were taken from the posterior distributions of the parameters via a Markov Chain Monte Carlo (MCMC) algorithm. This was performed using JAGS software \citep{plummer2003jags}. Standard diagnostic checks using trace plots and the Gelman and Rubin diagnostic were used to check convergence \citep{gelman2013bayesian}.

Best estimates of all parameters of interest were taken to be the median of the relevant posterior samples. The 95\% Bayesian credible intervals were calculated by finding the 2.5\% and 97.5\% quantiles of the posterior samples. 

All code and data are available on GitHub: https://github.com/MJAlexander/fb-migration-bayes

\section{Results}
\begin{figure}[h!]
\centering
    \begin{subfigure}[b]{0.8\textwidth}
        \includegraphics[width=\textwidth]{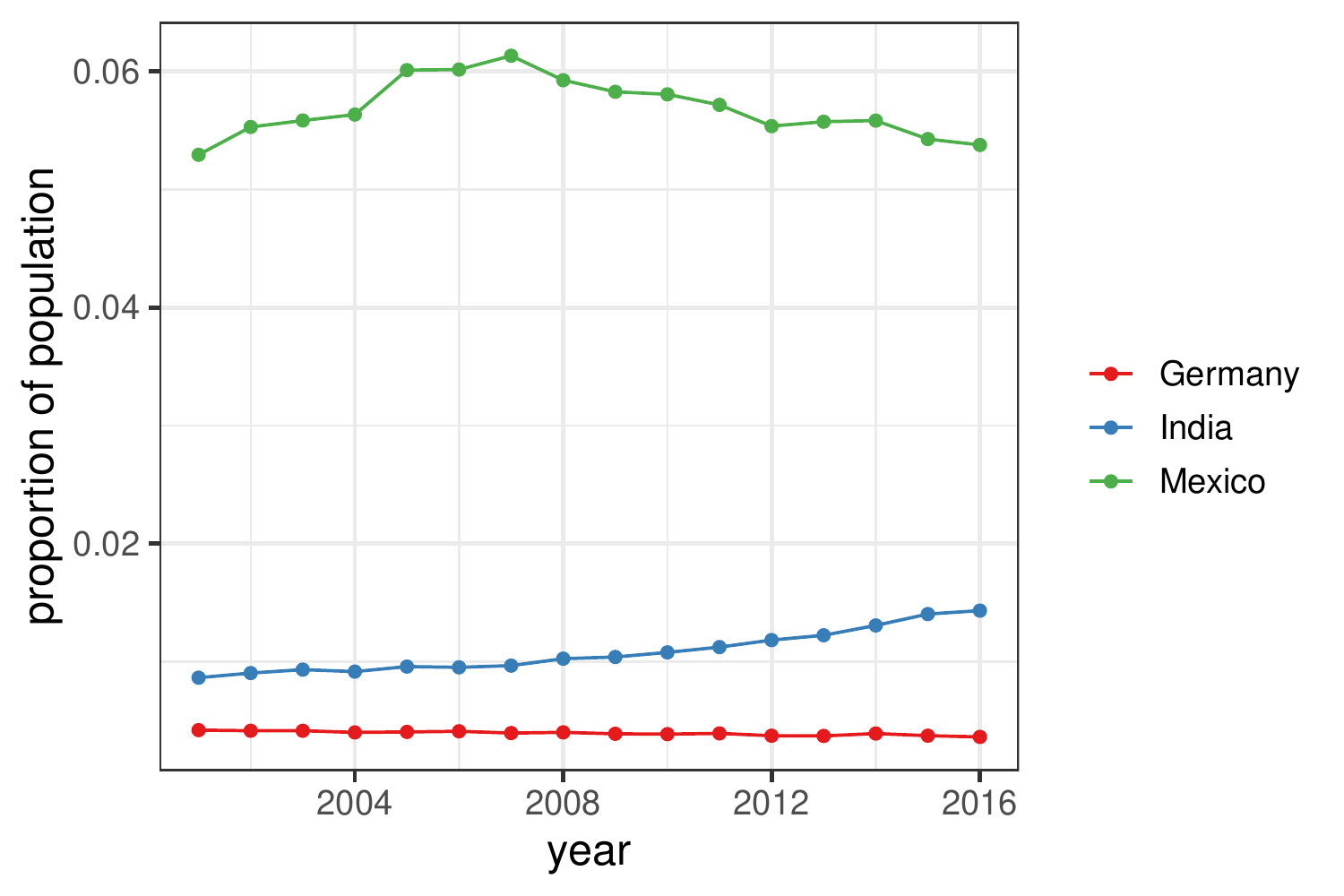}
        \caption{By proportion of the total population}
        \label{fig_props}
    \end{subfigure}
    \begin{subfigure}[b]{0.8\textwidth}
        \includegraphics[width=\textwidth]{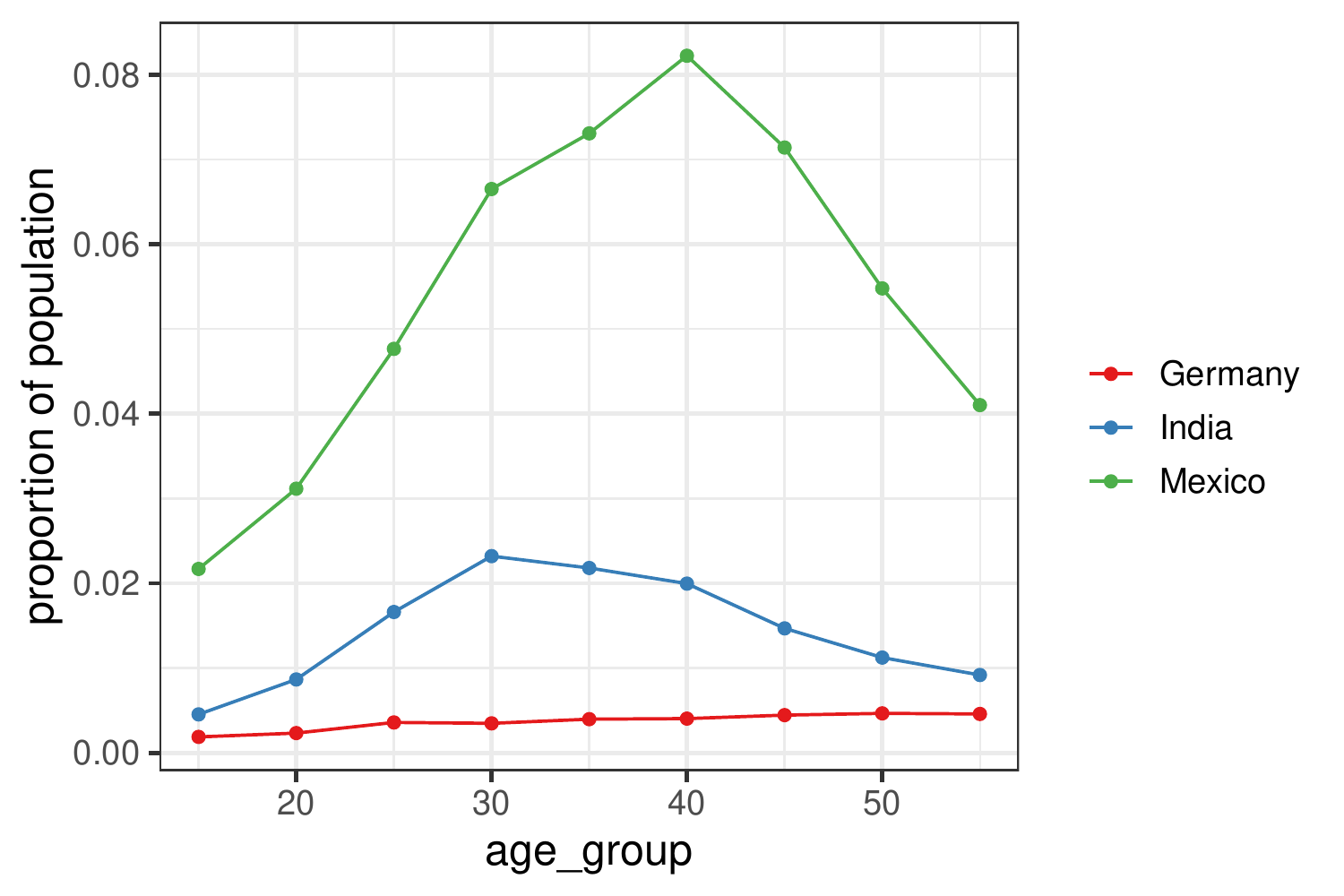}
        \caption{By proportion of each age group, 2016}
        \label{fig_age_dists}
    \end{subfigure}
\caption{German, Indian and Mexican migrants}
\end{figure}

We illustrate the model on male migrants from three different countries: Mexico, India, and Germany. These three migrant groups represent three different scenarios of levels and trends over time, as illustrated by the trends in the ACS data shown in Figures \ref{fig_props} and \ref{fig_age_dists}. 

Firstly, Mexican migrants make up a relatively large share of the overall population, but the proportion has been generally declining since around 2007. The age distribution at the national level peaks in the 40-44 year old age group. Secondly, Indian migrants make up a moderate proportion of the total population, but this share is increasing over time. The age distribution peaks at younger ages (30-34), compared to Mexicans. Finally, German migrants make up a low and declining share of the population. In contrast to the other migrant groups, the age distribution of German migrants at the national level is relatively flat, increasing slightly across age. 

\subsection{Bias adjustment of Facebook data}
\begin{figure}[h!]
\centering
\includegraphics[width=1\textwidth]{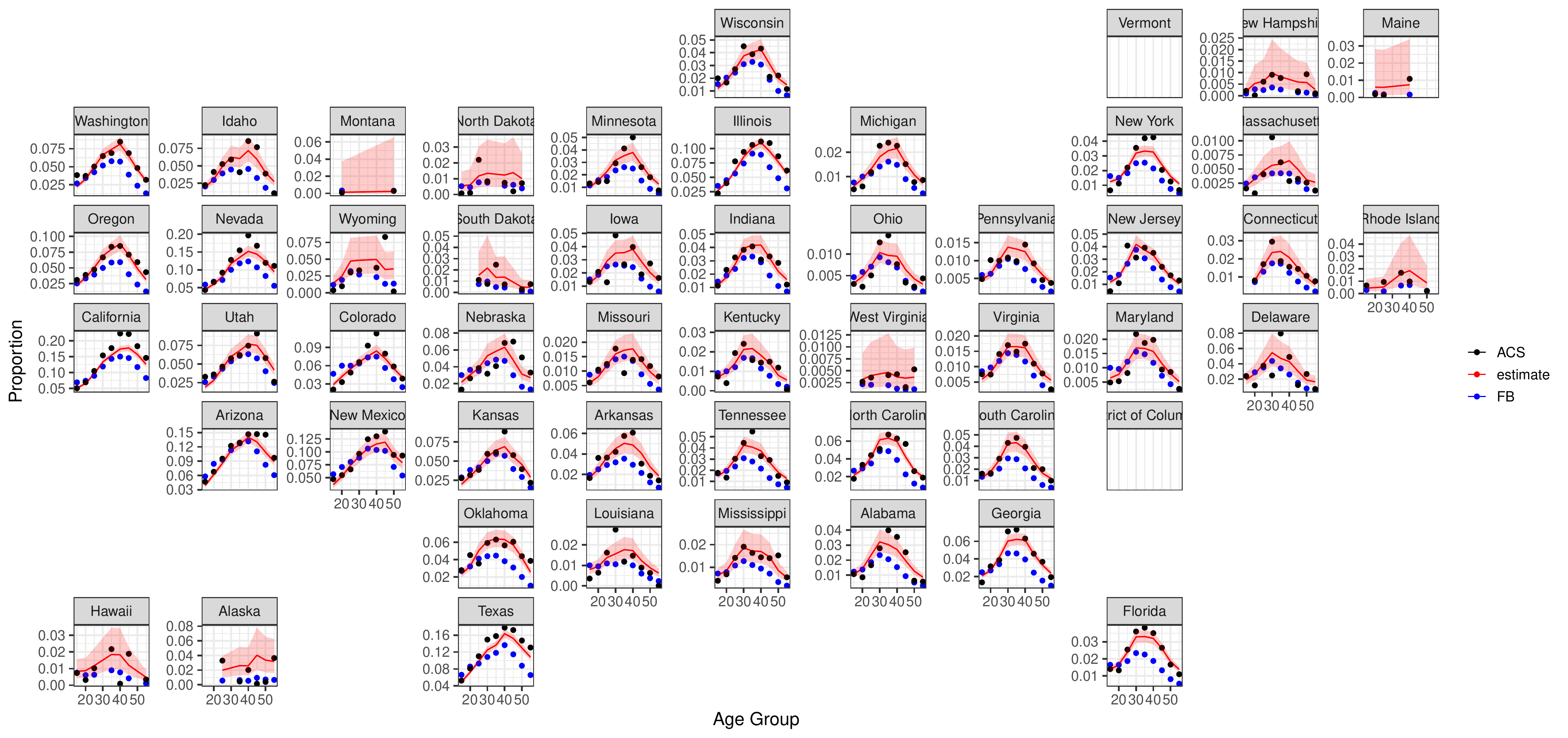}
\caption{Bias adjustment of Facebook data for Mexican migrants}
\label{fig_bias_MEX}
\end{figure}

We firstly illustrate the results of the bias-adjustment step of the Facebook data. Figure \ref{fig_bias_MEX} shows, for each US state and five-year age group where data are available, the proportion of migrants in each age group for the ACS data in 2016 (black dots), the un-adjusted Facebook data (blue dots), and the estimated bias-adjusted Facebook data (red line and associated shaded area) for Mexican migrants. Similar plots for migrants from India and Germany are shown in Appendix \ref{appendix_bias_adjustment}. The interpretation is that if the bias-adjustment step is working reasonably well, the red line would be close to the black dots. In general, this appears to be the case. 

For all three migrant groups, the raw Facebook data are generally lower than the ACS data, but the bias-adjustment model adjusts these values upwards. In general, across the three migrant groups and across states, the shape of the age distributions in the Facebook and ACS data are similar, with more substantial under-representation in Facebook in the older age groups. These systematic differences mean that the model works well to adjust the raw Facebook data based on age and state effects. 

\subsection{Nowcasts by age group and state}

\begin{figure}[ht]
\centering
\includegraphics[width=1\textwidth]{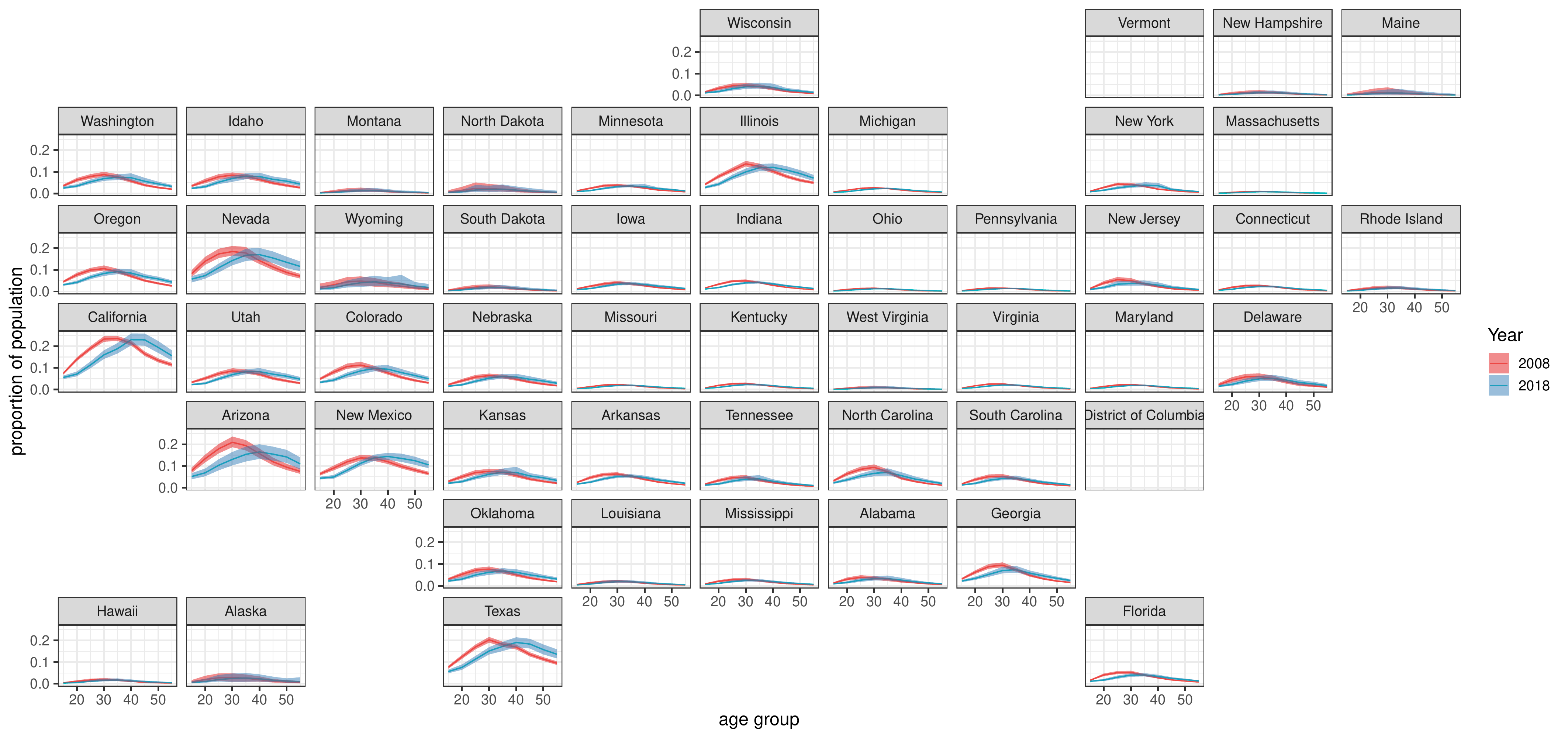}
\caption{Estimated and projected age distributions of Mexican migrants by state, 2008 and 2018}
\label{fig_MEX_state_age}
\end{figure}

Now we move on to short term projections by age and state. Figure \ref{fig_MEX_state_age} shows the estimated age distribution in 2008 (red) and projected distribution in 2018 (blue) for Mexican migrants. Similar plots for India and Germany can be found in Appendix \ref{appendix_age_facets}. 

For Mexico (Figure \ref{fig_MEX_state_age}), the relatively high proportions in the border states and on the West coast are apparent, with the highest proportions in California, Texas, Nevada and Arizona. Additionally, the age distribution of Mexican migrants is generally aging over time (shifting to the right), which is consistent with relatively constant stocks.

\subsubsection{Projected time series}

\begin{figure}
    \centering
    \begin{subfigure}[b]{0.8\textwidth}
        \includegraphics[width=\textwidth]{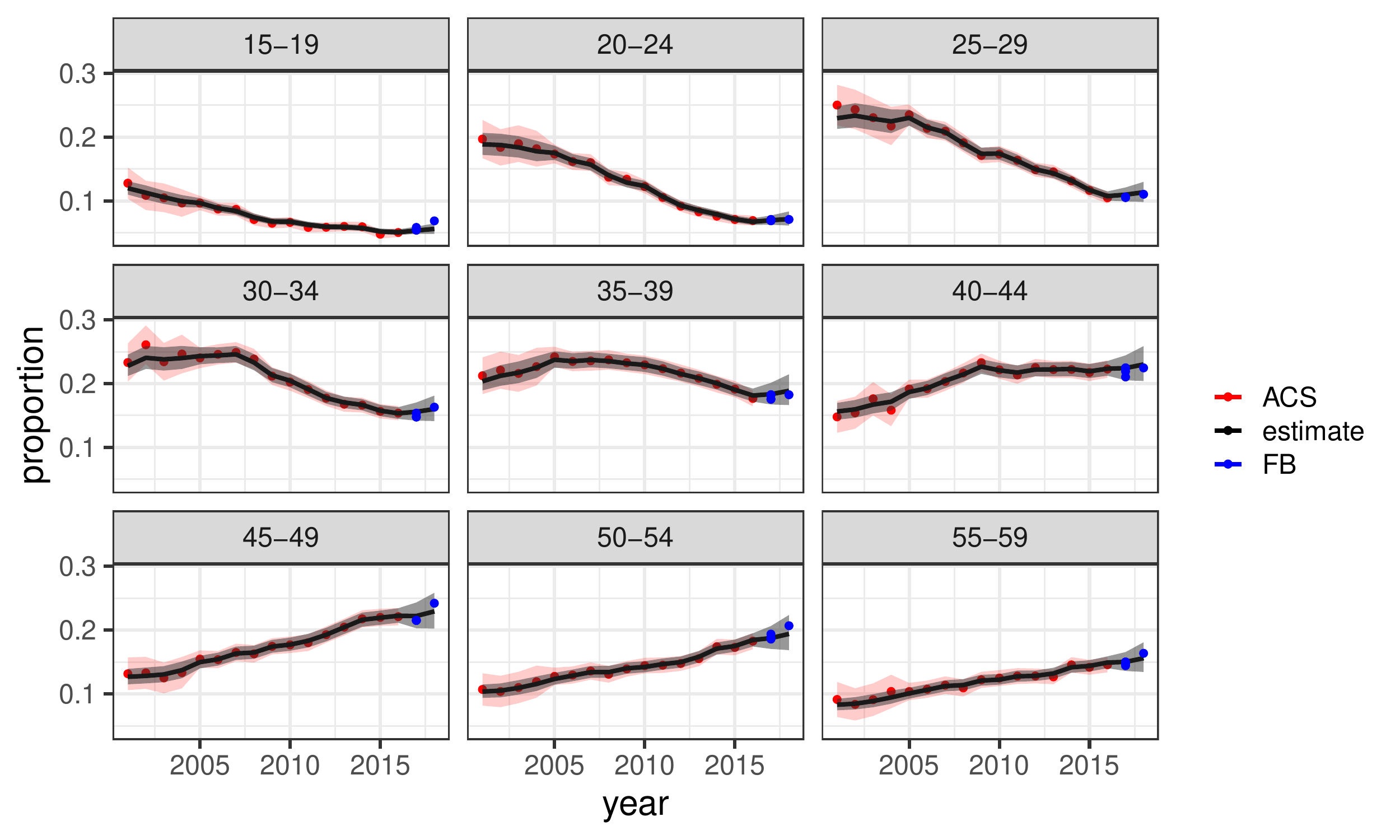}
        \caption{California}
        \label{fig:MEX_CA}
    \end{subfigure}
    \begin{subfigure}[b]{0.8\textwidth}
        \includegraphics[width=\textwidth]{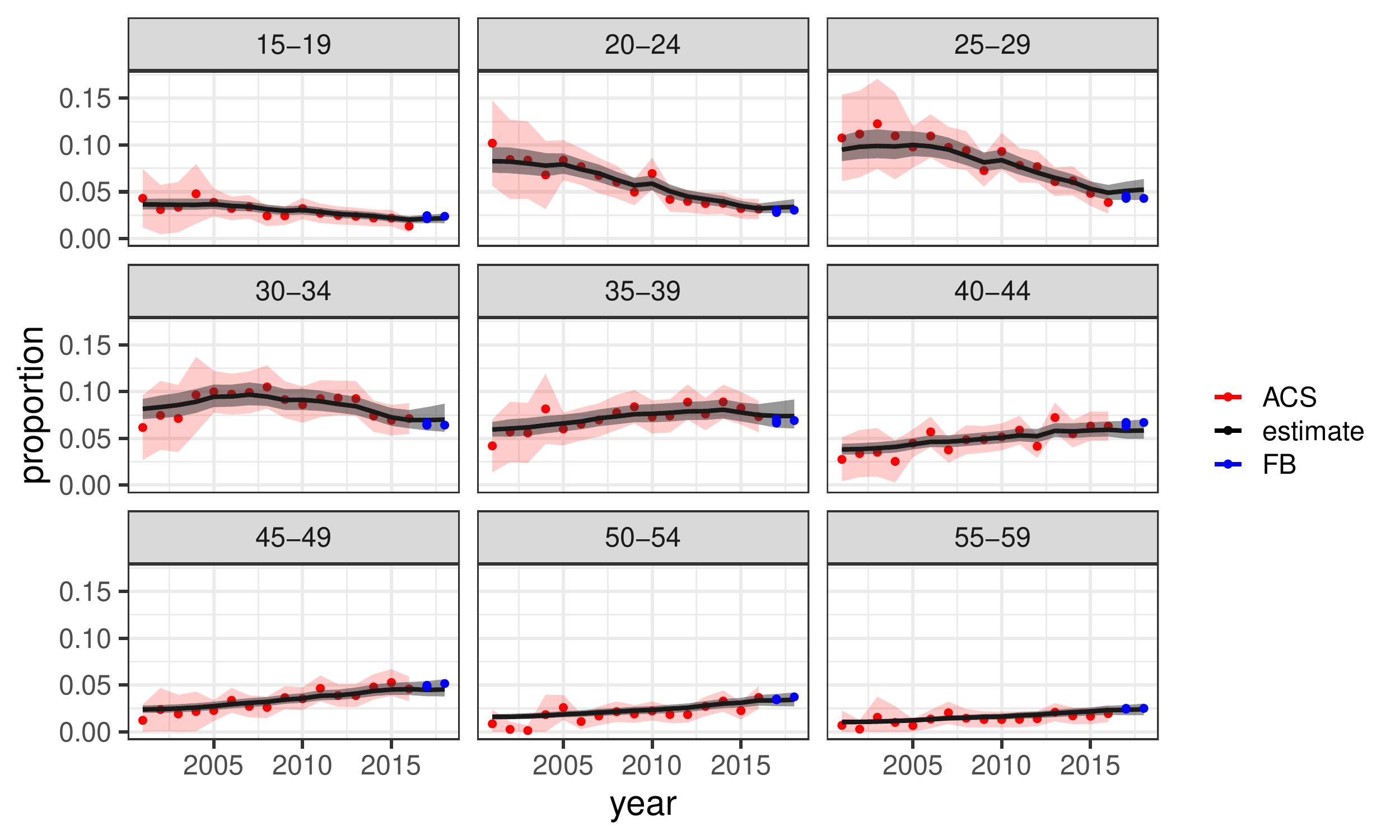}
        \caption{Georgia}
        \label{fig:MEX_GA}
    \end{subfigure}
    \caption{Mexican male migrants by age group, California and Georgia, 2001-2018}\label{fig:MEX_ts}
\end{figure}

\begin{figure}
    \centering
    \begin{subfigure}[b]{0.8\textwidth}
        \includegraphics[width=\textwidth]{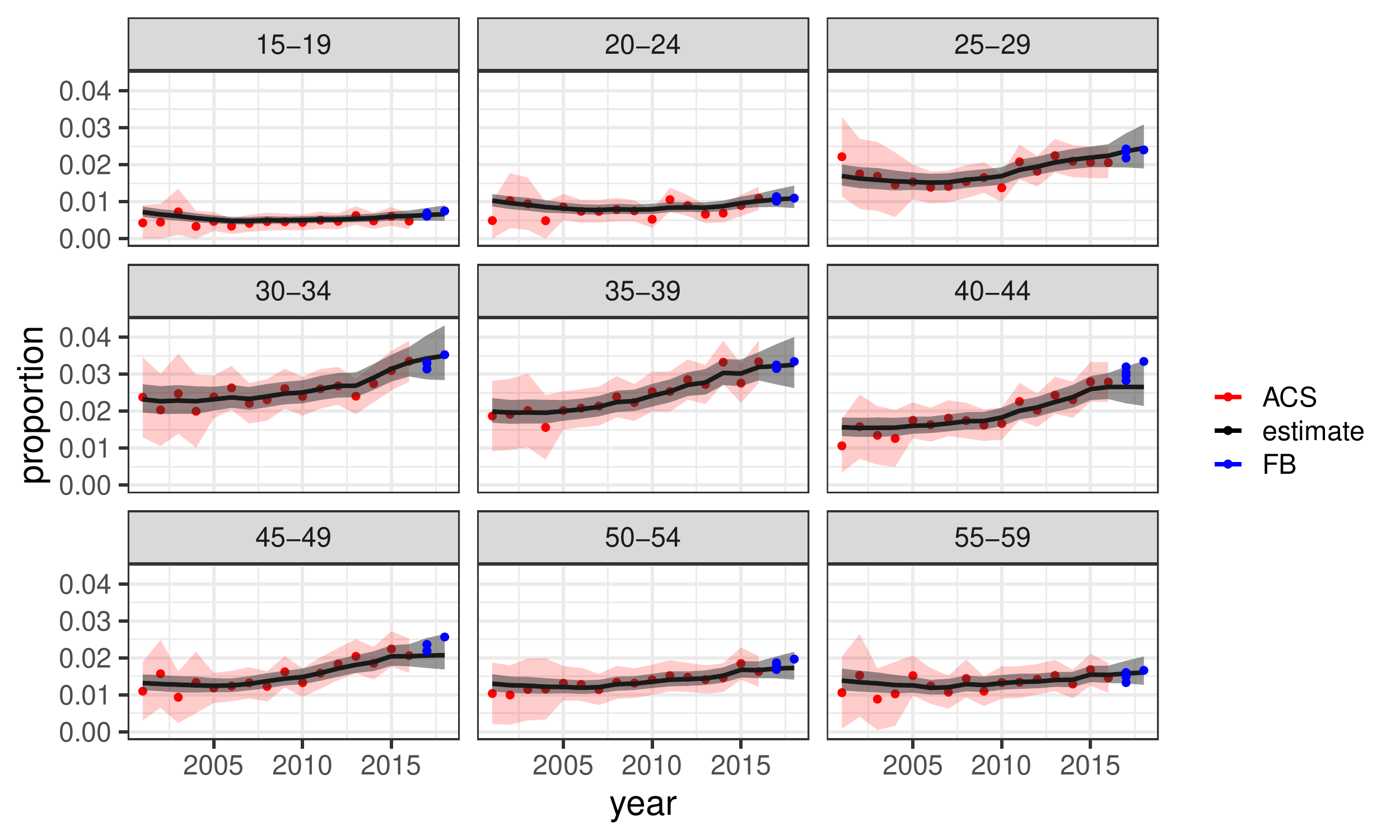}
        \caption{California}
    \end{subfigure}
    \begin{subfigure}[b]{0.8\textwidth}
        \includegraphics[width=\textwidth]{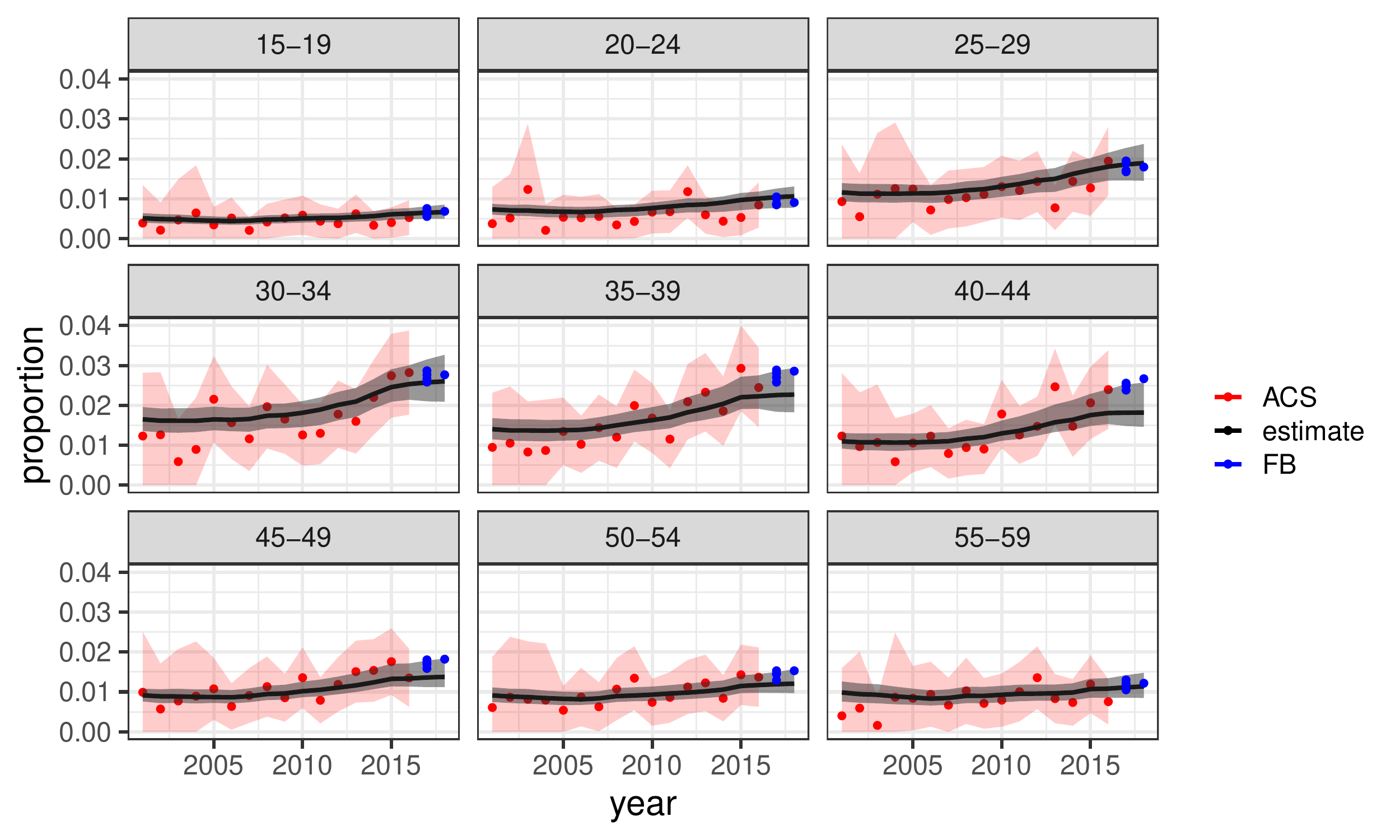}
        \caption{Georgia}
    \end{subfigure}
    \caption{Indian male migrants by age group, California and Georgia, 2001-2018}\label{fig:IND_ts}
\end{figure}

\begin{figure}
    \centering
    \begin{subfigure}[b]{0.8\textwidth}
        \includegraphics[width=\textwidth]{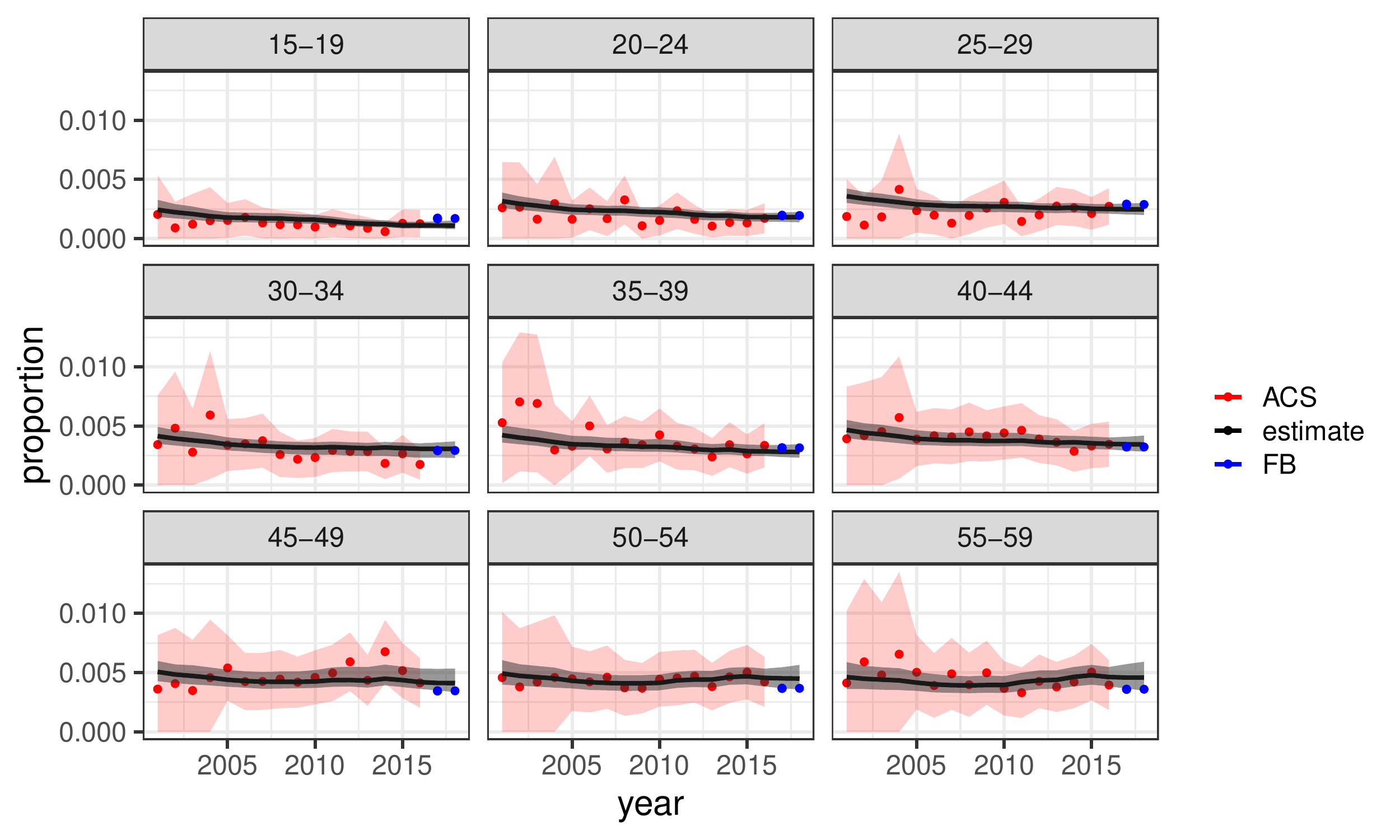}
        \caption{California}
    \end{subfigure}
    \begin{subfigure}[b]{0.8\textwidth}
        \includegraphics[width=\textwidth]{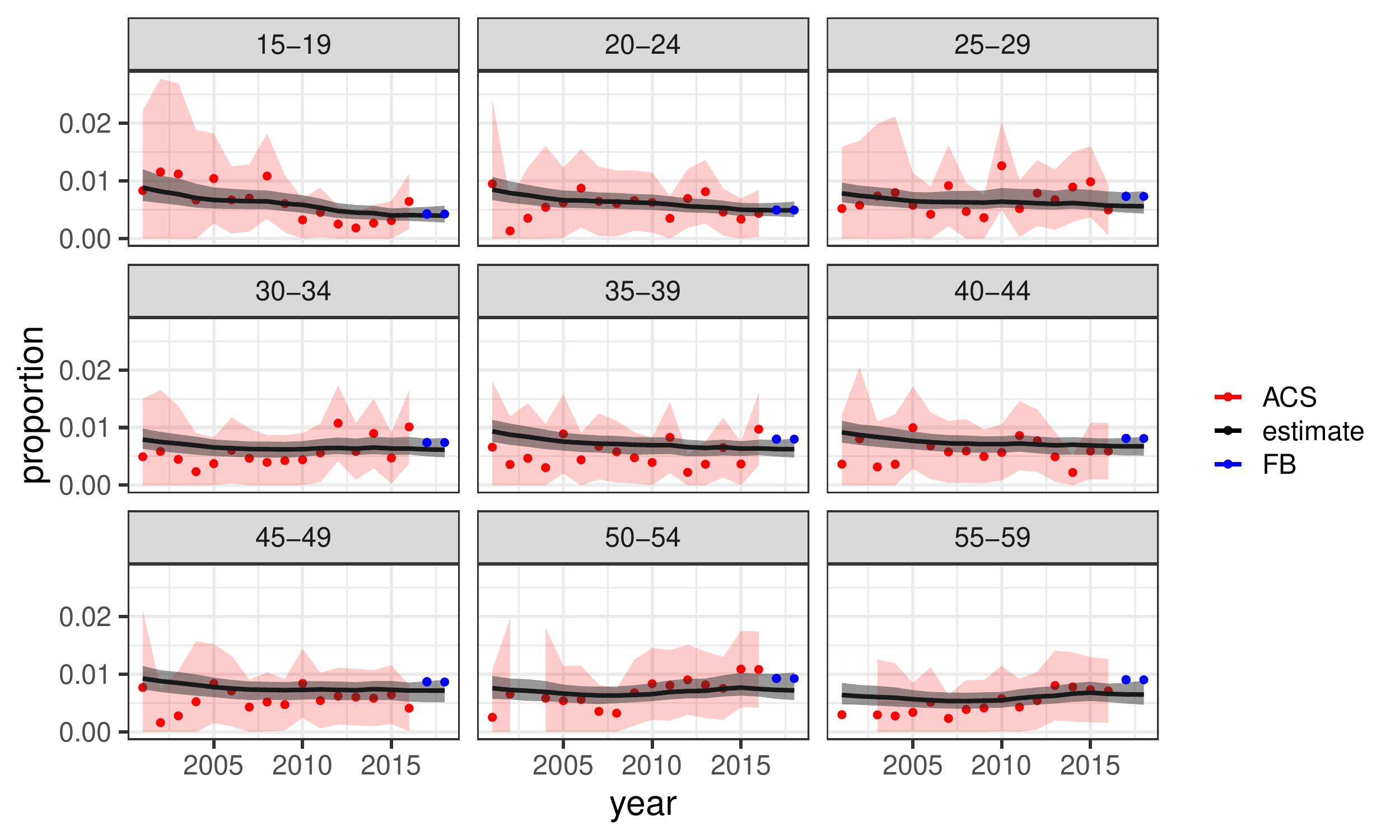}
        \caption{Georgia}
    \end{subfigure}
    \caption{German male migrants by age group, California and Georgia, 2001-2018}\label{fig:DEU_ts}
\end{figure}

Figures \ref{fig:MEX_ts} to \ref{fig:DEU_ts} zoom in on two states for each migrant group and show how the Facebook data are used to project forward the time series to the most recent two years (2017 and 2018). The full estimated and projected time series from 2001 to 2018 is shown. In the figures, each facet is a five-year age group. The red dots and associated shaded area represent the ACS data and sampling standard errors; these data are broadly available from 2001 to 2016, although some observations are missing (if sample sizes in the ACS were too small to capture information about migrants in that particular state and age group). The blue dots represent the (adjusted) Facebook observations, which are available in years 2017 and 2018. The black line and associated shaded area is the model estimate and 95\% uncertainty intervals. 

Mexican males in California (Figure \ref{fig:MEX_CA}) represent by far the highest proportions of any of the migrant origin/ state combinations considered. The proportion is as high as 0.25 in some age groups, for example 25-29 year olds in 2001 and 40-44 year olds in 2018. As a consequence, the sampling error around the ACS data for this migrant group is relatively small and the model estimates closely follow these data. For the most recent two years, where only Facebook data are available, note that the model estimates do not follow the data as closely and the uncertainty around the model estimates increases. This reflects the fact that there are more sources of error associated with the Facebook data.  

In Georgia (Figure \ref{fig:MEX_GA}) the levels of Mexican migrants are around half as high as in California. Due to smaller sample sizes in Georgia, the standard errors around the ACS data are much larger, and as such the model estimates do not follow the data as closely. However, the trends for Mexican migrants in California, and Georgia are broadly the same: decreases in the younger age groups, and increases in the older age groups, representing an aging stock of migrants. 

For Indian males in California and Georgia (Figure \ref{fig:IND_ts}), the proportions are much lower than for the Mexican migrant population, peaking at around 3-4\% of the population in the 30-39 year old age groups. The proportions are increasing over time, however, particularly in the 25-44 age bracket. Finally, for German male migrants in California and Georgia (Figure \ref{fig:DEU_ts}), we see low and constant migrant proportions. The uncertainty around the ACS data is already relatively high, and so there is not so much of an increase in uncertainty in the final two years. 

%%%%%%%%%%%%%%%%%%%%%%%%%%%%%%%%%%%%%%%%%%%%%%
\clearpage
\newpage
\section{Validation}\label{validation}

We evaluated the performance of the Bayesian model compared to other reasonable forecasting alternatives. To do this, we ran the model on data from 2001 to 2016, and forecast migration stocks in 2017. We then compared these forecasts to the actual ACS data in 2017. We compared the accuracy of the Bayesian model forecast to forecasts produced by three other models:

\begin{enumerate}
\item \textbf{Three-year moving average} of the ACS data. This is one of the simplest options available and does not require the Facebook data or any statistical modeling. 
\item \textbf{Facebook data only}: Estimates are based just on the available Facebook data in 2017, after it has been adjusted for biases. 
\item \textbf{ACS time series model}: Here, we ran the Bayesian hierarchical time series model described in section \ref{model} above, but just using data from the ACS (no Facebook). 
\end{enumerate}

In order to assess model performance, we compare the root mean squared error (RMSE):

\begin{equation}
RMSE = \sqrt{\frac{\sum_n \left( \hat{p}_{g, 2017} - p^{ACS}_{g 2017}\right)^2}{N}}
\end{equation}

where $\hat{p}_{g, 2017}$ is the estimated proportion of migrants from a particular group $g$, $ p^{ACS}_{g 2017}$ is the equivalent proportion from the ACS and $N$ is the size of the group. Here, the $g$ can refer to any combination of age group, state and migrant origin. 

Table \ref{table:rsme} shows the overall RMSE for the four models for Mexican, Indian and German migrants. The main result is  that in each of the three migrant groups, the Bayesian model presented (which combines the ACS and bias-adjusted Facebook data and thus is referred to as the 'combined model'), produces the lowest RMSE and thus the most accurate forecasts. The overall results also illustrate that the Bayesian hierarchical time series model produces substantially more accurate forecasts compared to a simple moving average or the bias-adjusted Facebook data alone, producing RMSEs that are up to an order of magnitude smaller. This gain in accuracy is much larger than the gain moving from ACS-only to the combined model, although there is still a gain in each case. 

\begin{table}[h!]
\centering
\begin{tabular}{lrrr}
                                       & \multicolumn{1}{l}{}                  & \multicolumn{1}{l}{}                  & \multicolumn{1}{l}{}        \\ \hline
\multicolumn{1}{l|}{Model}             & \multicolumn{1}{l|}{Mexico}           & \multicolumn{1}{l|}{India}            & \multicolumn{1}{l}{Germany} \\ \hline
\multicolumn{1}{l|}{Moving average}    & \multicolumn{1}{r|}{0.01280}          & \multicolumn{1}{r|}{0.0480}           & 0.0129                      \\
\multicolumn{1}{l|}{Facebook}          & \multicolumn{1}{r|}{0.01210}          & \multicolumn{1}{r|}{0.00584}          & 0.0142                      \\
\multicolumn{1}{l|}{ACS}               & \multicolumn{1}{r|}{0.00995}          & \multicolumn{1}{r|}{0.00453}          & 0.00263                     \\
\multicolumn{1}{l|}{\textbf{Combined}} & \multicolumn{1}{r|}{\textbf{0.00970}} & \multicolumn{1}{r|}{\textbf{0.00356}} & \textbf{0.00261} \\
\hline
\end{tabular}
\caption{Overall RMSE by model and migrant origin}
\label{table:rsme}
\end{table}

\begin{figure}
    \centering
    \begin{subfigure}[b]{0.5\textwidth}
        \includegraphics[width=\textwidth]{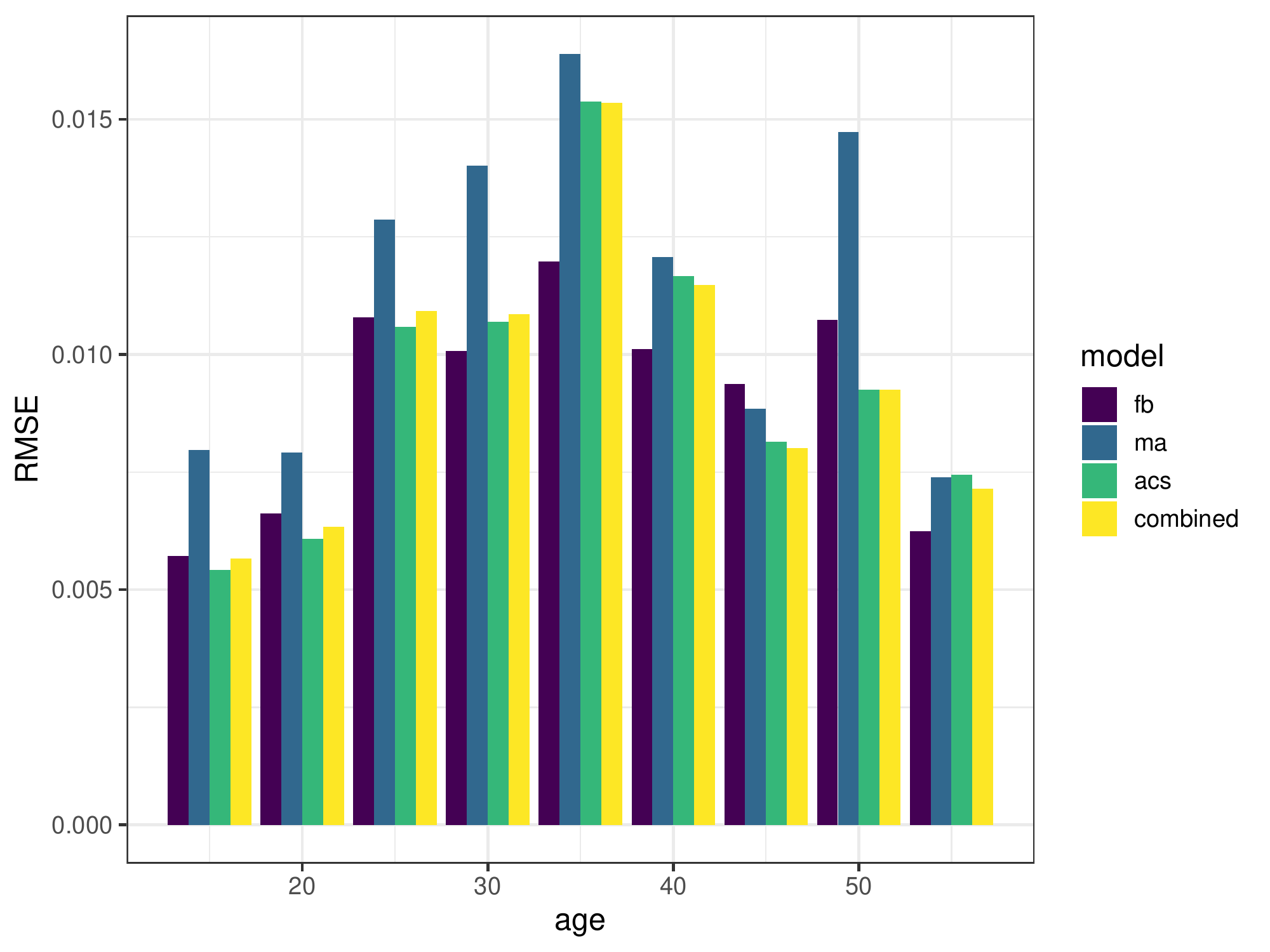}
        \caption{Mexico}
    \end{subfigure}
    \begin{subfigure}[b]{0.5\textwidth}
        \includegraphics[width=\textwidth]{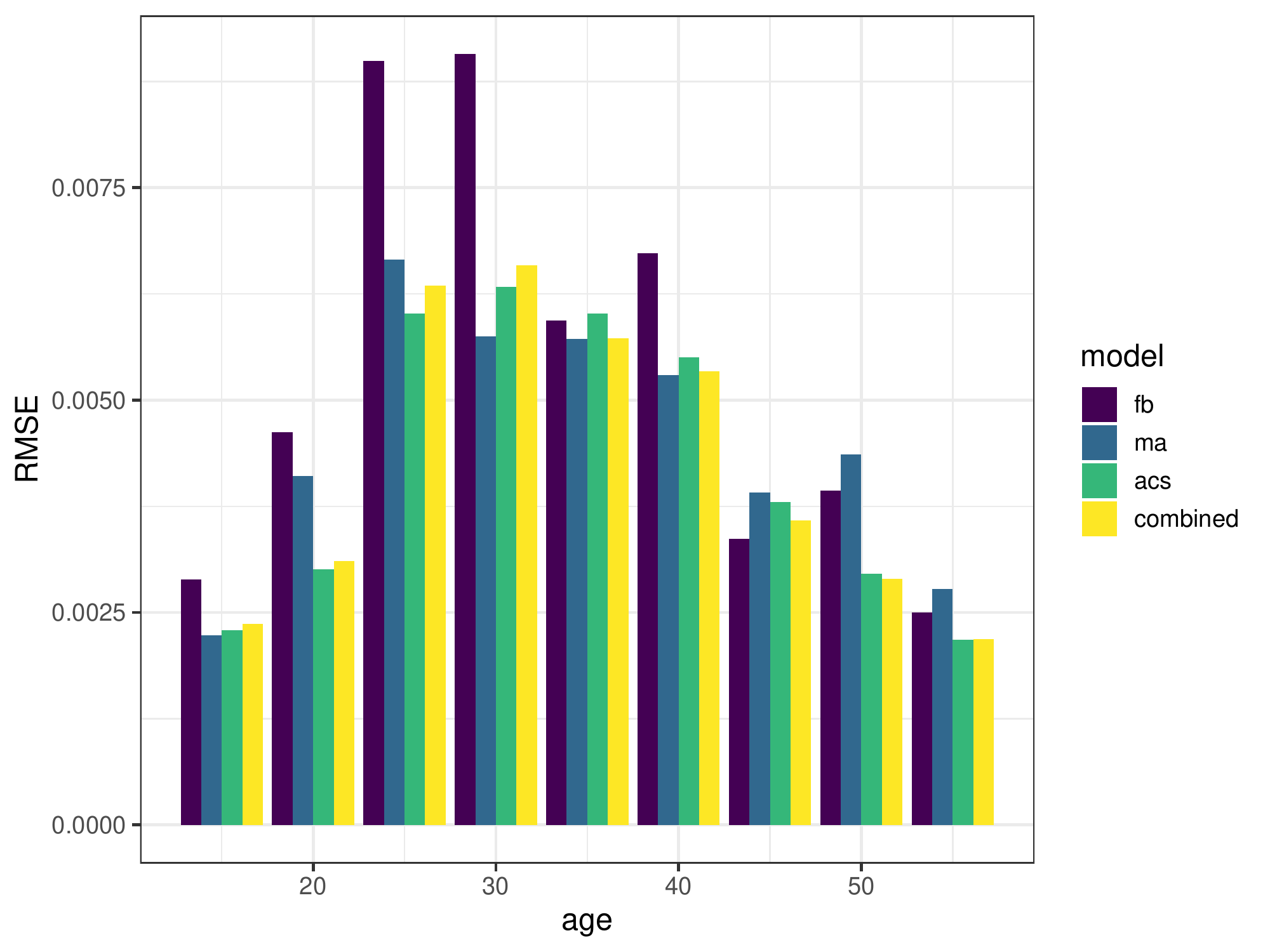}
        \caption{India}
    \end{subfigure}
        \begin{subfigure}[b]{0.5\textwidth}
        \includegraphics[width=\textwidth]{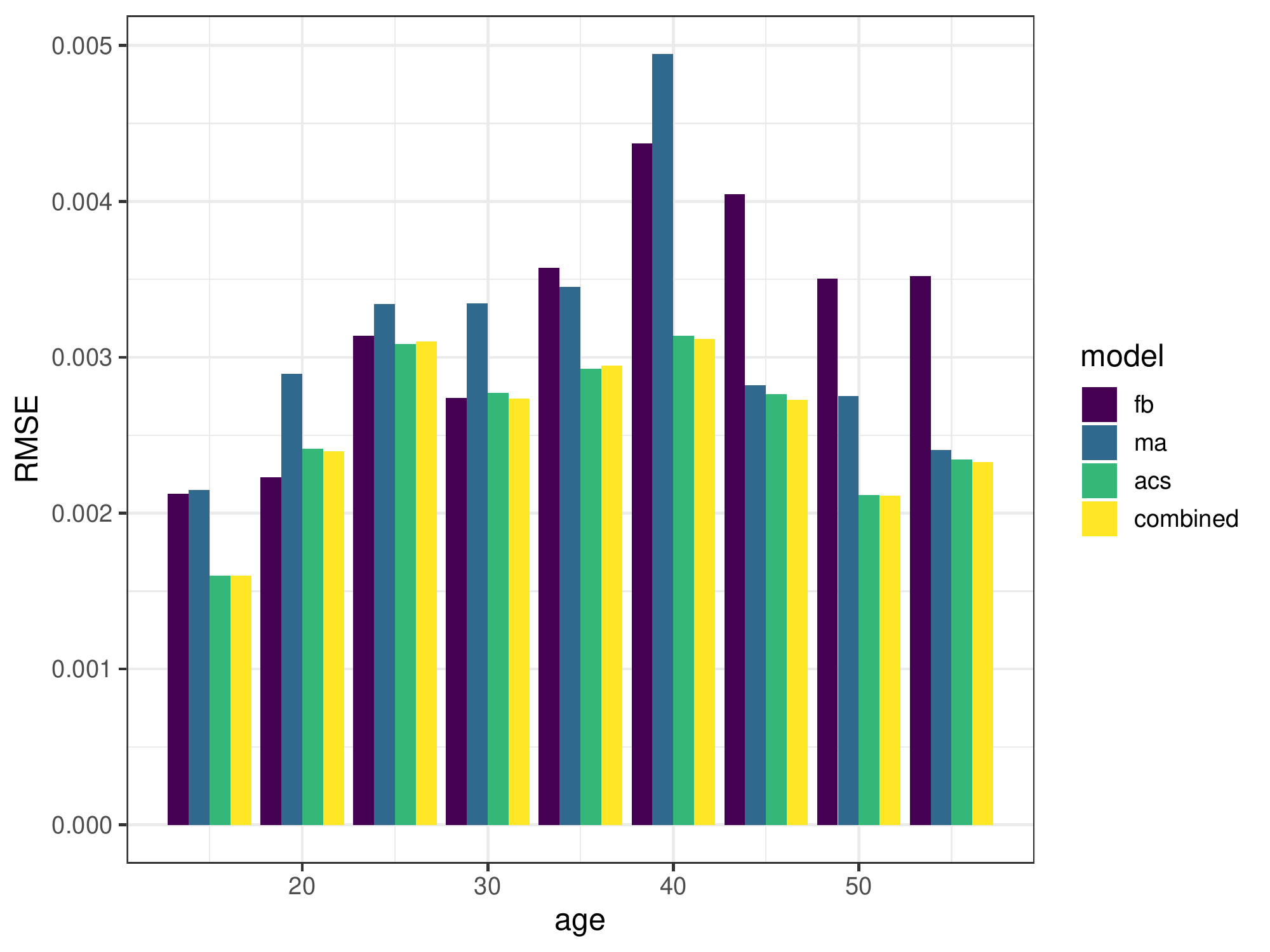}
        \caption{Germany}
    \end{subfigure}
    \caption{RMSE by age group and model}\label{fig:rsme_age}
\end{figure}

Figure \ref{fig:rsme_age} illustrates the RMSE by age group and model type for each of the three migrant groups. Similar plots by state can be found in Appendix \ref{appendix_validation}. For Mexico, there is generally an incremental decline in the RMSE moving from the Facebook-only model, to moving average, to ACS, to the combined model. The RMSE is highest in the 30-34 year old age group, which is also where the proportion of migrants is highest (see Figure \ref{fig:MEX_ts}). For India (Figure \ref{fig:rsme_age}b)), the RMSE is particularly high from the Facebook-only model in the 25-34 age bracket. As a consequence, the combined model RMSEs in those age groups are higher in the combined model versus the ACS model. For Germany, the gain in accuracy in moving to the hierarchical time series set-up is much larger. This is most likely related to the fact that the proportions of German migrants are in general a lot lower than for Mexico or India, and so there are noticeable gains in pooling information across state, age and time. 

To summarize, the validation exercise comparing the one-year-out predictions from a range of models to the migrant proportions reported in the ACS illustrate both (i) the strength of the proposed Bayesian hierarchical time series model as a general framework, and (ii) the additional information obtained from including up-to-date Facebook data compared to just historical ACS data alone. 

%%%%%%%%%%%%%%%%%%%%%%%%%%%%%%%%%%%%%%%%%%%%%%%%%%
\newpage
\section{Discussion}\label{discussion}

As the size and frequency of migration movements continues to increase worldwide, new sources of data are being considered in order to better understand both historical and future migration trends. There is a growing body of work considering the feasibility of using social media data to achieve such goals, from platforms such as Facebook, Twitter and LinkedIn. While the granularity of social media use varies widely (from individual geo-tagged tweets, to aggregated advertising demographic data, as is in this paper), the common challenges of using such data remain: firstly, to adequately adjust for known biases in the social media data, primarily as a consequence of the non-representativeness of the population of social media users; and secondly, to meaningfully combine information from social media data with information from more traditional data sources, such as surveys or censuses. 

In this paper we presented a statistical framework to achieve these goals in the context of producing short-term projections of migrant stocks in the United States. The model includes a bias-adjustment process of the Facebook data, and a `principal components time series' model, which allows for the projection of trends in stocks into the future, considering both Facebook and ACS data. The model allows for different types of uncertainty around the different data sources, and shares information on migration trends over time and pools across geographic space. Illustrative results were presented for three separate migrant groups: Mexicans, Indians and Germans. The results of the validation exercise, comparing projections with 2017 ACS data, suggest that the proposed model improves prediction of short-term trends when compared to viable alternatives. 

The validation exercise illustrated the substantial gain in accuracy achieved when moving to the Bayesian hierarchical time series model, regardless of whether or not the Facebook data were included. While the benefits of including the Facebook data in this particular case were relatively marginal, more generally the Facebook data has the advantage of being up-to-date and essentially available in real time. Thus, in a situation of a `shock', such as a natural disaster or other event, the collection of Facebook data allows for a more immediate estimate of the effects of that shock on migration. The combination of these data with past trends allows for the identification of surprising increases or decreases, that are out of the expected bounds based on historical patterns. 

There are several limitations of the proposed model, which naturally lead into avenues for future work. Firstly, the bias-adjustment model assumes that the systematic bias in the Facebook data (by age and state) is constant over time. In reality, it is reasonable to believe that the biases in the Facebook data are changing over time, as the composition of the underlying Facebook population changes. The relationship between the age/location composition of the Facebook and the actual population (as measured by the ACS) could be investigated in future work. Secondly, the bias-adjustment model also assumes that the non-sampling error is constant over Facebook's `waves' of data collection --- that is, sources of error that include changes in how the population of reach is calculated, or other computational reasons, are assumed to be constant. In practice, and in other work using these data (Alexander et al. 2019), we have observed that this is probably not the case, and needs to be further investigated to better understand non-migration-related fluctuations over time. While we only consider two data sources in this paper, the general statistical framework could easily be extended to include information from other sources. Future work could also include taking advantage of the rich demographic and socioeconomic data available through the Facebook Advertising Platform, including information on education and occupation. 

While this work focused on a model for the estimation of migrant stocks, the philosophy of combining social media data with more traditional data sources in  one statistical framework -- allowing for different sources of uncertainty -- can be readily extended to model other demographic indicators. Indeed, the underlying time series model was itself an extension of principal component techniques that have been previously been used in demography to study mortality and fertility. We show the strength of combining more traditional demographic modeling techniques and data with newer sources of data to gain insights into underlying population processes.

%%%%%%%%%%%%%%%%%%%%%%
\newpage
\appendix

\section{Bias adjustment plots for migrants from India and Germany} \label{appendix_bias_adjustment}

\begin{figure}[h!]
\centering
\includegraphics[width=1\textwidth]{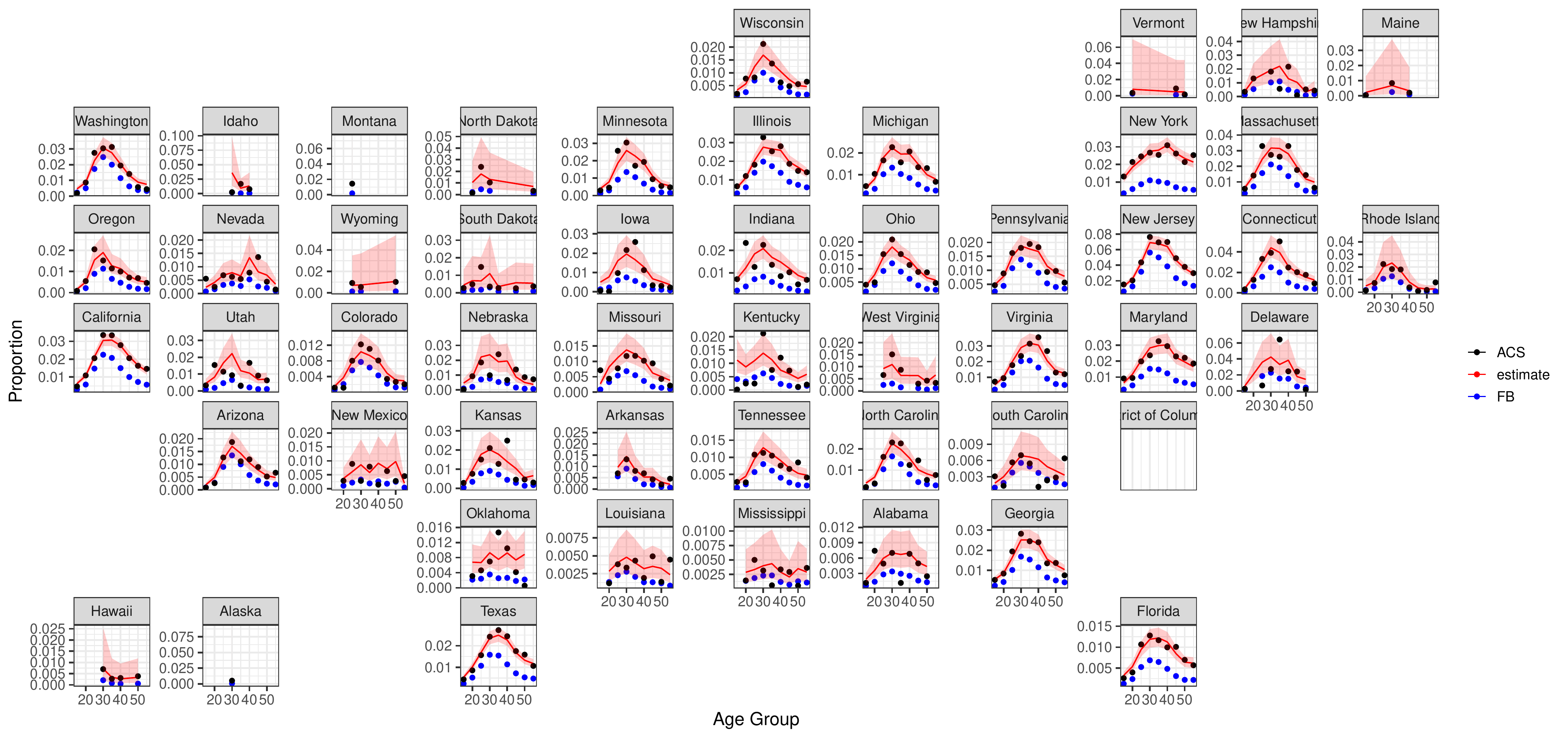}
\caption{Bias adjustment of Facebook data for Indian migrants}
\label{fig_bias_IND}
\end{figure}

\begin{figure}[h!]
\centering
\includegraphics[width=1\textwidth]{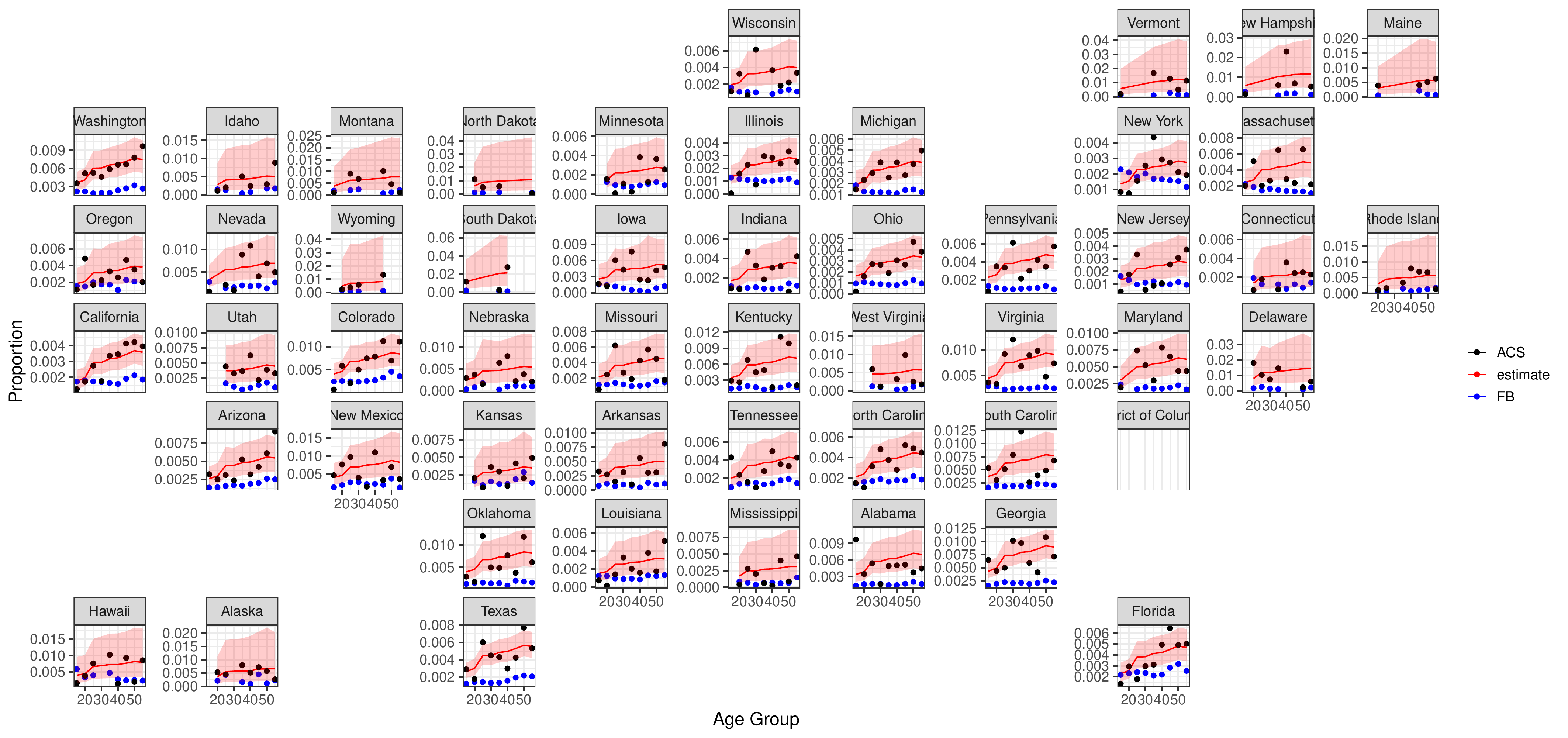}
\caption{Bias adjustment of Facebook data for German migrants}
\label{fig_bias_DEU}
\end{figure}

%%%%%%%%%%%%%%%%%%%%%%%%%
\newpage
\section{Age distributions in 2008 and 2018 for Indian and Germany migrants by state}

\begin{figure}[ht]
\centering
\includegraphics[width=1\textwidth]{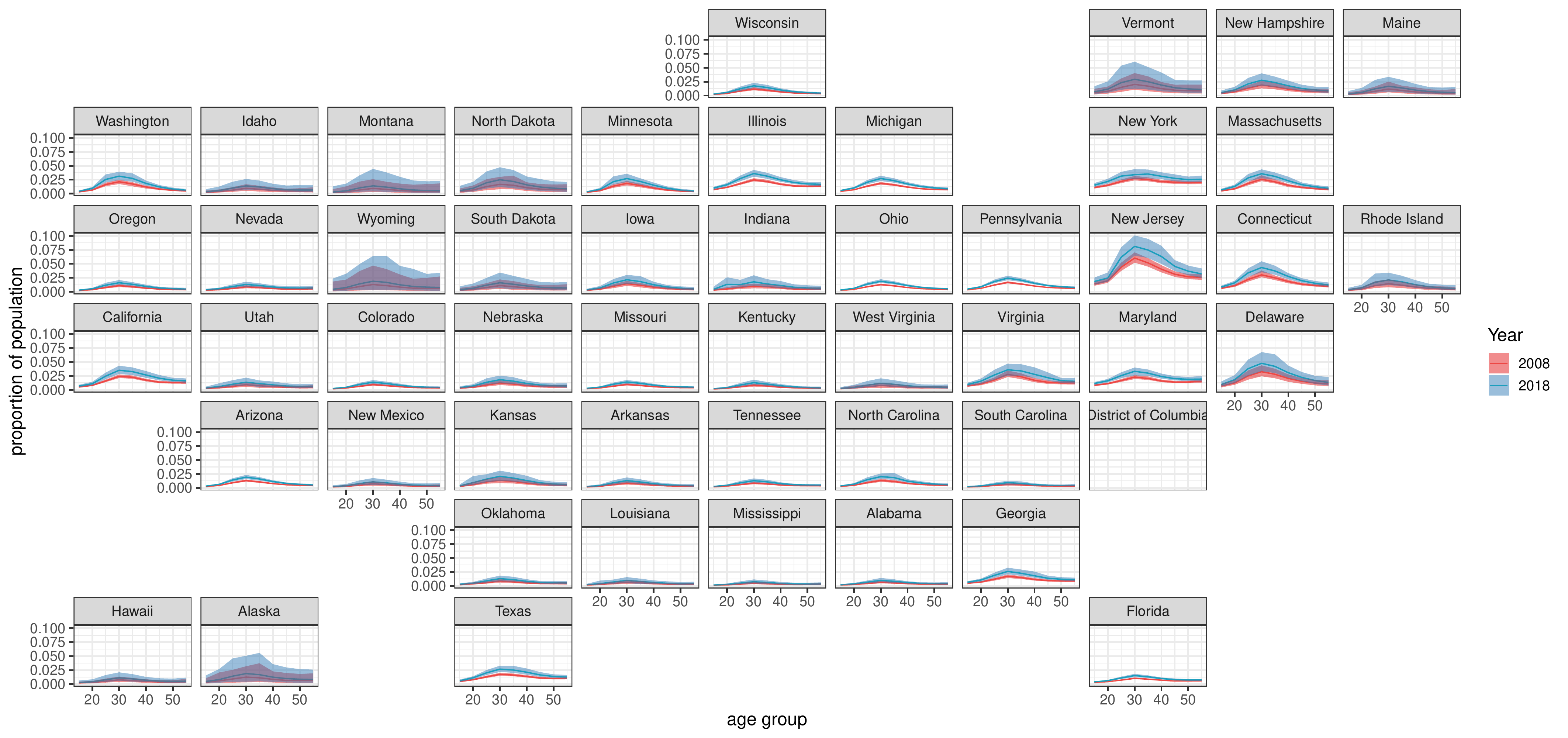}
\caption{Estimated and projected age distributions of Indian migrants by state, 2008 and 2018}
\label{fig_IND_state_age}
\end{figure}

\begin{figure}[ht]
\centering
\includegraphics[width=1\textwidth]{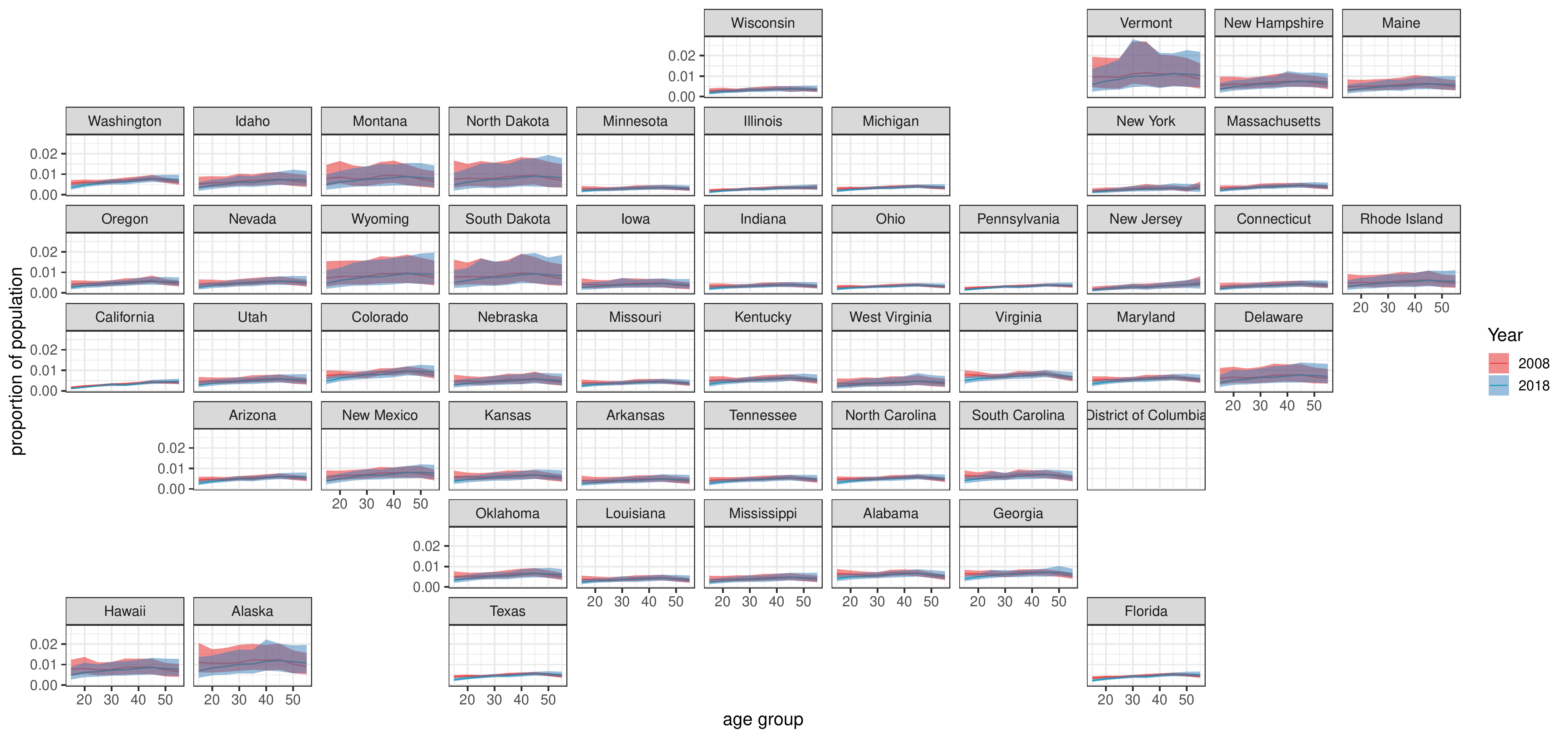}
\caption{Estimated and projected age distributions of German migrants by state, 2008 and 2018}
\label{fig_DEU_state_age}
\end{figure}

For India (Figure \ref{fig_IND_state_age}) we see that the proportions across age groups have generally increased over the decade, with relatively high proportions concentrated in the northeast region of the country. Unlike Mexico, the age distribution is fairly constant, with the highest proportions generally being in the 30-34 age group. The German male migrant populations (Figure \ref{fig_DEU_state_age}) by state show relatively flat, low and unchanging levels over the decade 2008--2018. 

%%%%%%%%%%%%
\clearpage
\newpage
\section{Validation results by state} \label{appendix_validation}

\begin{figure}[ht]
\centering
\includegraphics[width=1\textwidth]{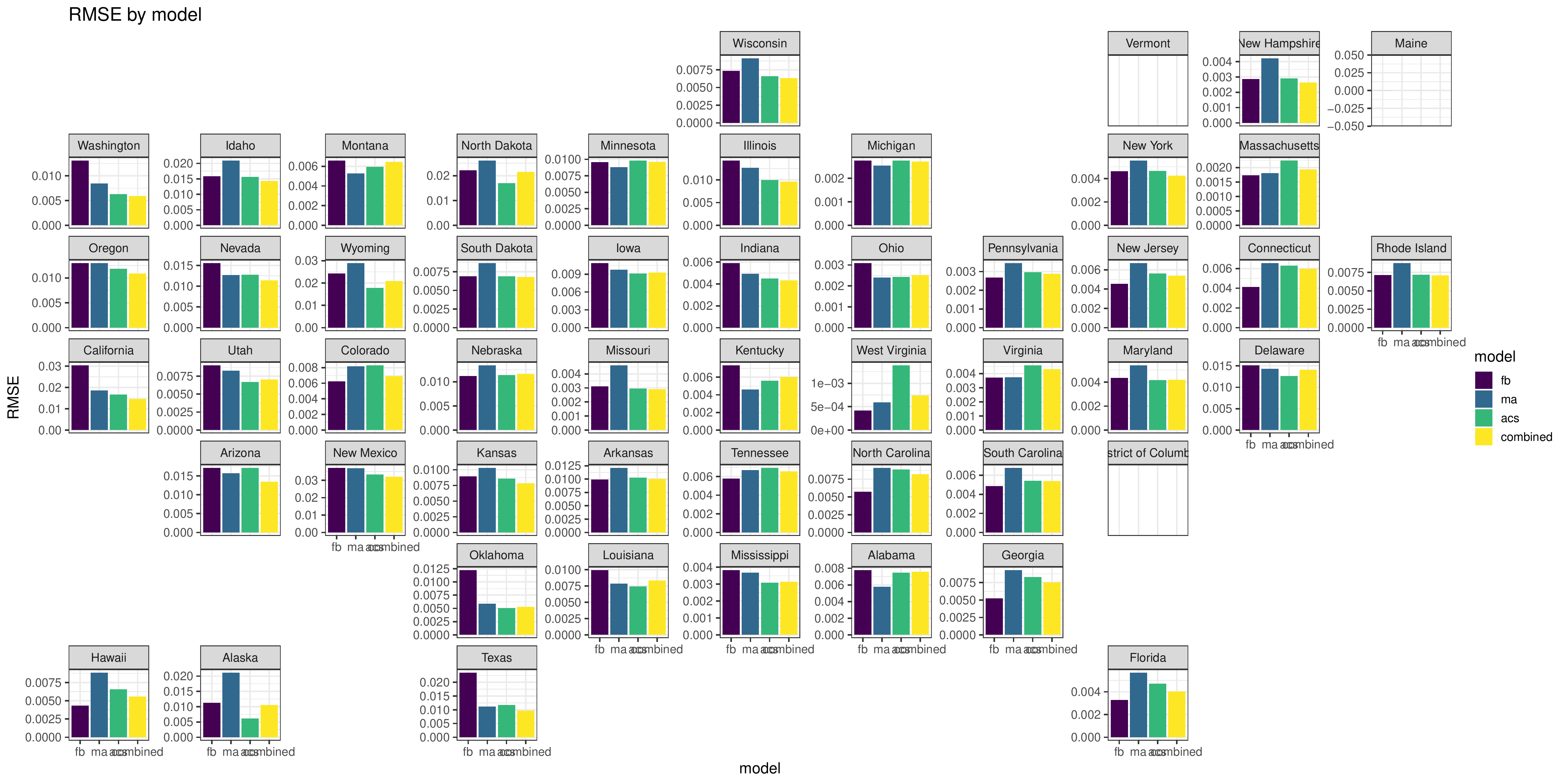}
\caption{RSME by state and model type for Mexican migrants in 2018}
\label{fig_MEX_state_rsme}
\end{figure}

\begin{figure}[ht]
\centering
\includegraphics[width=1\textwidth]{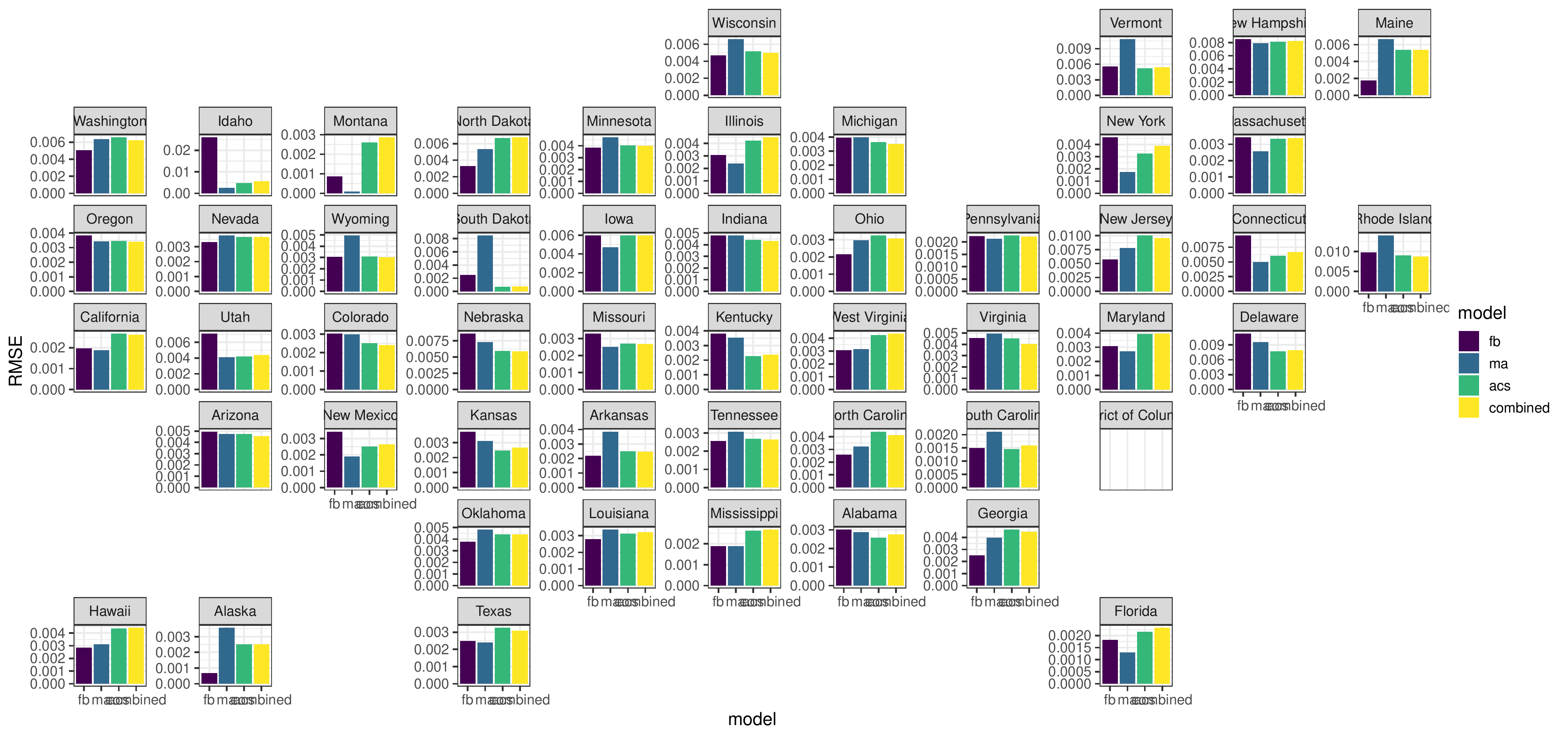}
\caption{RSME by state and model type for Indian migrants in 2018}
\label{fig_IND_state_rsme}
\end{figure}

\begin{figure}[ht]
\centering
\includegraphics[width=1\textwidth]{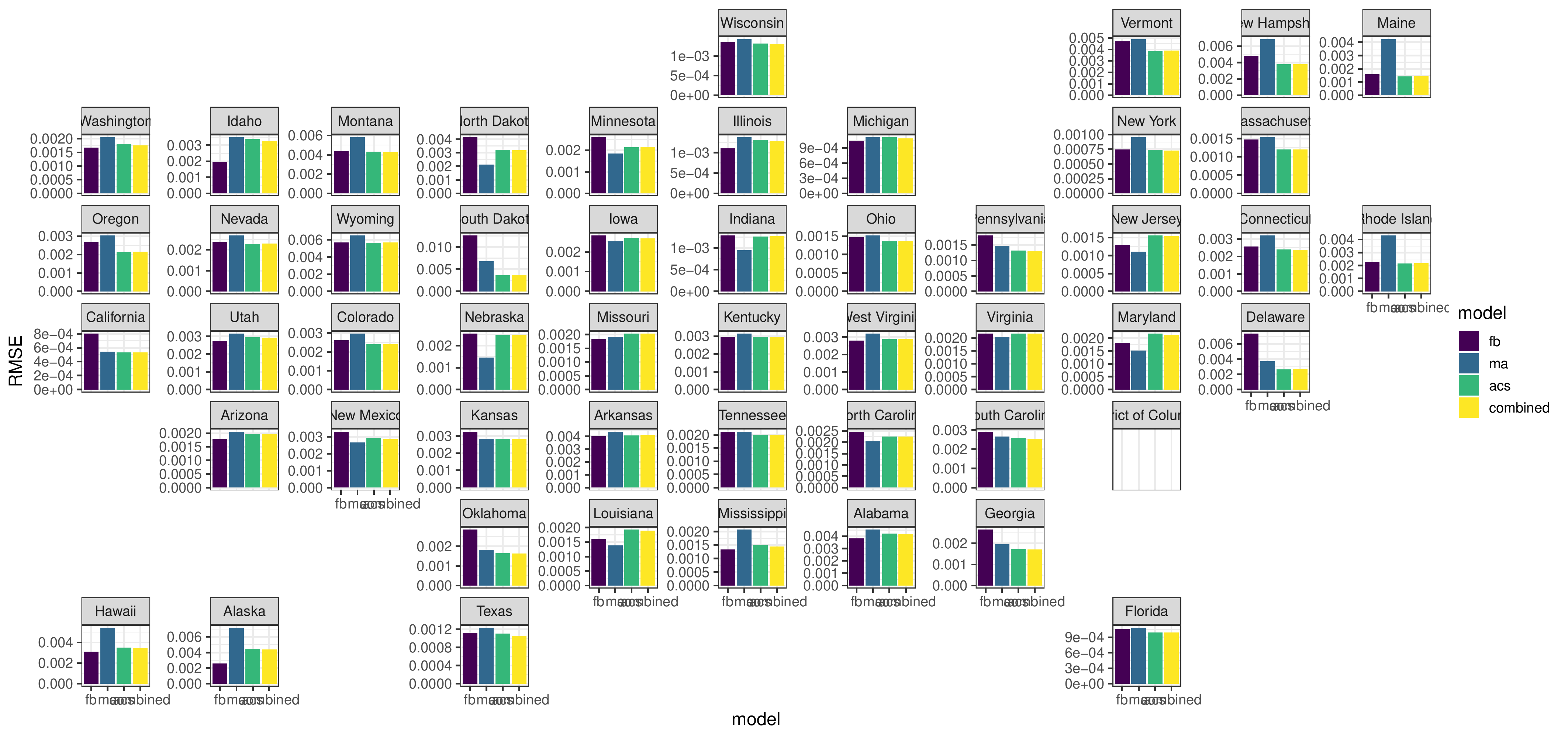}
\caption{RSME by state and model type for German migrants in 2018}
\label{fig_DEU_state_rsme}
\end{figure}

The figures below show the RMSE for each model, by state of destination in the US, for Mexican, Indian and German migrants. The figures show that, in general, the combined model out-performs the other three alternative models. For Mexico (Figure \ref{fig_MEX_state_rsme}), this is particularly the case for the states with relatively high proportions of Mexican migrants. Results are more variable for Indian migrants (Figure \ref{fig_IND_state_rsme}), with the combined model performing relatively poorly in the West coast states but well in the mid-West states. For German migrants (Figure \ref{fig_DEU_state_rsme}), the geographic pooling appears to vastly improve the projection accuracy. 

\clearpage
\newpage
\bibliographystyle{chicago}
\bibliography{fb_refs}
\end{document}